\documentclass[3p,11pt,authoryear]{elsarticle}

\usepackage{amsfonts}
\usepackage{amsmath}
\usepackage{amssymb}
\usepackage{amsthm}
\usepackage{mathrsfs}
\usepackage{graphics}
\usepackage{lscape}
\usepackage{changebar}
\usepackage{graphicx}
\usepackage{epsfig}
\usepackage{bm}
\usepackage{color}

\newcommand{\apc}{\alpha\!_{_{PC}}}
\newcommand{\ald}{\alpha\!_{_{LD}}}

\newcommand{\B}[1]{{\bm{#1}}}
\newcommand{\C}[1]{{\mathcal{#1}}}
\newcommand{\pa}{\partial}

\journal{Journal of the Mechanics and Physics of Solids}

\begin{document}

\begin{frontmatter}

\title{Dynamic instabilities of frictional sliding at a bimaterial interface}

\author{Efim A. Brener$^1$, Marc Weikamp$^2$, Robert Spatschek$^2$, Yohai Bar-Sinai$^3$ and Eran Bouchbinder$^3$}
\address{$^1$ Peter Gr\"unberg Institut, Forschungszentrum J\"ulich, D-52425 J\"ulich, Germany \\
$^2$ Max-Planck-Institut f\"ur Eisenforschung GmbH, D-40237 D\"usseldorf, Germany\\
$^3$ Chemical Physics Department, Weizmann Institute of Science, Rehovot 7610001, Israel}
\date{\today}
\begin{abstract}
Understanding the dynamic stability of bodies in frictional contact steadily sliding one over the other is of basic interest in various disciplines such as physics,
solid mechanics, materials science and geophysics. Here we report on a two-dimensional linear stability analysis of a deformable solid of a finite height $H$,
steadily sliding on top of a rigid solid within a generic rate-and-state friction type constitutive framework, fully accounting for elastodynamic effects.
We derive the linear stability spectrum, quantifying the interplay between stabilization related to the frictional constitutive law and destabilization related both
to the elastodynamic bi-material coupling between normal stress variations and interfacial slip, and to finite size effects.
The stabilizing effects related to the frictional constitutive law include velocity-strengthening friction (i.e.~an increase in frictional resistance with increasing slip velocity, both instantaneous and under steady-state conditions) and a regularized response to normal stress variations.
We first consider the small wave-number $k$ limit and demonstrate that homogeneous sliding in this case is universally unstable, independently of the details of the friction law.  This universal instability is mediated by propagating waveguide-like modes, whose fastest growing mode is characterized by a wave-number satisfying $k H\!\sim\!{\C O}(1)$ and by a growth rate that scales with $H^{-1}$. We then consider the limit $k H\!\to\!\infty$ and derive the stability phase diagram in this case.
We show that the dominant instability mode travels at nearly the dilatational wave-speed in the opposite direction to the sliding direction.
In a certain parameter range this instability is manifested through unstable modes at all wave-numbers, yet the frictional response is shown to be mathematically well-posed.
Instability modes which travel at nearly the shear wave-speed in the sliding direction also exist in some range of physical parameters.
Previous results obtained in the quasi-static regime appear relevant only within a narrow region of the parameter space. Finally, we show that a finite-time regularized response to normal stress variations, within the framework of generalized rate-and-state friction models, tends to promote stability. The relevance of our results to the rupture of bi-material interfaces is briefly discussed.
\end{abstract}


\end{frontmatter}


\section{Background and motivation}
\label{sec:intro}

The dynamic stability of steady-state homogeneous sliding between two macroscopic bodies in frictional contact is a basic problem of interest in various scientific disciplines such as physics, solid mechanics, materials science and geophysics. The emergence of instabilities may give rise to rich dynamics and play a dominant role in a broad range of frictional phenomena~\citep{Ruina1983, Ben-Zion2001, Scholz2002, Ben-Zion2008, Gerde2001, Ibrahim1994a, Ibrahim1994, DiBartolomeo2010, Tonazzi2013, Baillet2005, Behrendt2011, Meziane2007}. The response of a frictional system to spatiotemporal perturbations, and the accompanying instabilities, are governed by several physical properties and processes. Generally speaking, one can roughly distinguish between bulk effects (i.e.~the constitutive behavior and properties of the bodies of interest, their geometry and the external loadings applied to them) and interfacial effects related to the frictional constitutive behavior. The ultimate goal of a theory in this respect is to identify the relevant physical processes at play, to quantify the interplay between them through properly defined dimensionless parameters and to derive the stability phase diagram in terms of these parameters, together with the growth rate of various unstable modes.

As a background and motivation for what will follow, we would like first to briefly discuss the various players affecting the stability of frictional sliding, along with stating some relevant results available in the literature. Focusing first on bulk effects, it has been recognized that when considering isotropic linear elastic bodies, there is a qualitative difference between sliding along interfaces separating bodies made of identical materials and interfaces separating dissimilar materials. In the former case, there is no coupling between interfacial slip and normal stress variations, while in the latter case --- due to broken symmetry --- such coupling exists~\citep{Comninou1977a, Comninou1977b, Comninou1979, Weertman1980, Andrews1997, Ben-Zion1998, Adams2000, Cochard2000, Rice2001, Ranjith2001, Gerde2001, Adda-Bedia2003, Ampuero2008a}. This coupling may lead to a reduction in the normal stress at the interface and consequently to a reduction in the frictional resistance. Hence, bulk material contrast (i.e.~the existence of a bi-material interface) potentially plays an important destabilizing role in the stability of frictional sliding. Another class of bulk effects is related to the finite geometry of any realistic sliding bodies and the type of loading applied to them
(e.g. velocity or stress boundary conditions). To the best of our knowledge, the latter effects are significantly less explored in the literature (but see~\citet{Rice1983, Ranjith2009, Ranjith2014}).

In relation to interfacial effects, it has been established that sliding along a bi-material interface described by the classical Coulomb friction law, $\tau\!=\!\sigma f$ ($\tau$ is the local friction stress, $\sigma$ is the local compressive normal stress and $f$ is a constant friction coefficient), is unstable against perturbations at all wavelengths and irrespective of the value of the friction coefficient, when the bi-material contrast is such that the generalized Rayleigh wave exists~\citep{Ranjith2001}. The latter is an interfacial wave that propagates along frictionless bi-material interfaces, constrained not to feature opening~\citep{Weertman1963, Achenbach1967, Adams1998, Ranjith2001}. It is termed the generalized Rayleigh wave because it coincides with the ordinary Rayleigh wave when the materials are identical and it exists when the bi-material contrast is not too large. In fact, the response to perturbations in this case is mathematically ill-posed~\citep{Renardy1992, Adams1995, Martins1995a, Martins1995b, Simoes1998, Ranjith2001}. Ill-posedness, which is a stronger condition than instability (i.e.~all perturbation modes can be unstable, yet a problem can be mathematically well-posed), will be discussed later. It has been then shown that replacing Coulomb friction by a friction law in which the friction stress $\tau$ does not respond instantaneously to variations in the normal stress $\sigma$, but rather approaches $\tau\!=\!\sigma f$ over a finite time scale, can regularize the problem, making it mathematically well-posed~\citep{Ranjith2001}.

Subsequently, the problem has been addressed within the constitutive framework of rate-and-state friction models, where the friction stress depends both on the slip velocity and the structural state of the interface. Within this framework~\citep{Dieterich1978, Dieterich1979, Ruina1983, Rice1983, Heslot1994, Marone1998, Berthoud1999, Baumberger1999, Baumberger2006}, the simplest version of the friction stress takes the form $\tau\!=\!\sigma f(\phi,v)$,
where $v$ is the difference between the local interfacial slip velocities of the two sliding bodies and $\phi$
is a dynamic coarse-grained state variable\footnote{In principle there can be more than one internal state variables,
but we do not consider this possibility here.}. Under steady-state sliding conditions the state variable $\phi$
attains a steady-state value determined by $v$, $\phi_0(v)$. Within such a constitutive framework, the most relevant physical quantities for the question of stability, which will be extensively discussed below, are the instantaneous response to variations in the slip velocity, $\pa_v\!f(\phi, v)$ (the so-called ``direct effect''), and the variation of the steady-state frictional strength with the slip velocity, $d_v\!f(\phi_0(v),v)$~\citep{Rice2001}. Note that here and below we use the following shorthand notation: $d_v\!\equiv\!\tfrac{d}{dv}$ and $\pa_v\!\equiv\!\tfrac{\pa}{\pa{v}}$.

Previous studies have argued that an instantaneous strengthening response, i.e.~the experimentally well-established positive direct effect $\pa_v\!f(\phi, v)\!>\!0$ (which is associated with thermally activated rheology~\citep{Baumberger2006}), is sufficient to give rise to the existence of a quasi-static range of response to perturbations at sufficiently small slip velocities~\citep{Rice2001}. The existence of such a quasi-static regime is non-trivial (e.g. it does not exist for Coulomb friction); it implies that when very small slip velocities are of interest, one can reliably address the stability problem in the framework of quasi-static elasticity (i.e.~omitting inertial terms to begin with), rather than considering the full --- and more difficult --- elastodynamic problem and then take the quasi-static limit. Within such a quasi-static framework, it has been shown that $\pa_v\!f(\phi, v)\!>\!0$ can lead to stable response against sufficiently short wavelength perturbations, even if the interface is velocity-weakening in steady-state, $d_v\!f(\phi_0(v),v)\!<\!0$~\citep{Rice2001}. Furthermore, it has been shown that sufficiently strong velocity-strengthening, $d_v\!f(\phi_0(v),v)\!>\!0$, can overcome the destabilizing bi-material effect, leading to the stability of perturbations at all wavelengths in the quasi-static limit~\citep{Rice2001}.

Despite this progress, several important questions remain open. First, to the best of our knowledge the fully elastodynamic stability analysis of bi-material interfaces in the framework of rate-and-state friction models has not been performed. This is important since the quasi-static regime --- when it exists --- is expected to be valid only for very small slip velocities (as was argued in~\citet{Rice2001} and will be explicitly shown below). Second, a very recent compilation of a large set of experimental data for a broad range of materials~\citep{Bar-Sinai2014jgr} has revealed that dry frictional interfaces generically become velocity-strengthening over some range of slip velocities~\citep{Weeks1993, Marone1988, Marone1991, Kato2003, Shibazaki2003, BarSinai2012, Hawthorne2013, Bar-Sinai2013pre, Bar-Sinai2015SciRep}. In other cases, frictional interfaces are intrinsically velocity-strengthening~\citep{Perfettini2008, Noda2009, Ikari2009, Ikari2013}. As velocity-strengthening friction is expected to play a stabilizing role in the stability of frictional sliding, there emerges a basic question about the interplay between the stabilizing velocity-strengthening friction effect and the destabilizing bi-material effect, when elastodynamics is fully taken into account. Finally, in almost all of the previous studies we are aware of, the sliding bodies were assumed to be infinite (but see, for example, \cite{Rice1983, Ranjith2009, Ranjith2014}). Yet, realistic sliding bodies are of finite extent and the interaction with the boundaries may be of importance.

To address these issues we analyze in this paper the linear stability of a deformable solid of height $H$ steadily sliding on top of a rigid solid within a generic rate-and-state friction constitutive framework, fully taking into account elastodynamic effects. The rate-and-state friction constitutive framework includes a single state variable $\phi$, but is otherwise general in the sense that no special properties of $f(\phi,v)$ are being specified and rather generic dynamics of $\phi$ are considered. Nevertheless, we will be mostly interested in the physically relevant case in which the interface exhibits a positive instantaneous response to velocity changes, $\pa_v\!f(\phi, v)\!>\!0$ (positive direct effect)~\citep{Marone1998, Baumberger2006}, and is steady-state velocity-strengthening, $d_vf(\phi_0(v),v)\!>\!0$, over some range of slip velocities~\citep{Bar-Sinai2014jgr}. In addition, we will consider two variants of the constitutive model, each of which incorporates a regularized response to normal stress variations~\citep{Linker1992, Dieterich1992a, Prakash1992, Prakash1993, Prakash1998, Richardson1999, Bureau2000} .

While our analysis remains rather general, the main simplification we adopt is that we consider the limit of strong material
contrast, i.e.~we take one of the solids to be non-deformable.
The motivation for this is two-fold. First, we know that the bi-material effect that emerges from the
coupling between interfacial slip and normal stress variations becomes stronger as the material contrast increases~\citep{Rice2001}.
We are interested here in exploring the ultimate range of stability and consequently we focus on strongly dissimilar materials, which will allow us to extract upper bounds on the stability of bi-material frictional interfaces. Second, the strong dissimilar materials limit somewhat reduces the mathematical complexity involved and
makes the problem more amenable to analytic progress, as will be shown below. We suspect that this simplification does not imply qualitative differences compared to the finite material contrast case, though interesting quantitative differences may emerge and will be explored elsewhere. The finite material contrast case can be studied along the same lines, though it is more technically involved.

The structure of this paper is as follows; in Sect.~\ref{sec:ExecutiveSummary} the main results of the paper are listed. In Sect.~\ref{sec:math} the basic equations and the constitutive framework are introduced. In Sect.~\ref{sec:spectrum} the linear stability spectrum for finite height $H$ systems is derived, with a focus on standard rate-and-state friction. In Sect.~\ref{sec:finiteH} the linear stability spectrum is analyzed in the small wave-number $k$ limit, demonstrating the existence of a universal instability (independent of the details of the friction law) with a fastest growing mode characterized by a wave-number satisfying $k H\!\sim\!{\C O}(1)$ and a growth rate that scales with $H^{-1}$.
In Sect.~\ref{sec:infiniteH} the linear stability spectrum is analyzed in the large systems limit, $k H\!\to\!\infty$. We derive the stability phase diagram in terms of
a relevant set of dimensionless parameters that quantify the competing physical effects involved. We show that the dominant instability mode travels at nearly the dilatational
wave-speed (super-shear) in the opposite direction to the sliding motion. In a certain parameter range this instability is manifested through unstable modes at all wave-numbers,
yet the frictional response is shown to be mathematically well-posed. Instability modes which travel at nearly the shear wave-speed in the sliding direction are shown to exist
in a relatively small region of the parameter space. Finally, previous results obtained in the quasi-static regime~\citep{Rice2001} are shown to be relevant within a narrow region of the parameter space. In Sect.~\ref{sec:modified-models} a finite-time regularized response to normal stress variations, within the framework of generalized rate-and-state friction models, is studied. We show that this regularized response tends to promote stability. Section~\ref{sec:summary} offers a brief discussion and some concluding remarks.

\section{The main results of the paper}
\label{sec:ExecutiveSummary}

The analysis to be presented below is rather extensive and somewhat mathematically involved. Yet, we believe that it gives rise to a number of physically significant and non-trivial results. In order to highlight the logical structure of the analysis and its major outcomes, we list here the main results to be derived in detail later on:
\begin{itemize}
\item The stability of a deformable body of finite height $H$ steadily sliding on top of a rigid solid is studied. The linear stability spectrum is derived in the constitutive framework of velocity-strengthening rate-and-state friction models and an instantaneous response to normal stress variations. 
\item The spectrum takes the form of a complex-variable equation implicitly relating the real wave-number $k$ and the complex growth rate $\Lambda$ of interfacial perturbations. Physically, it represents the balance between the perturbation of the elastodynamic shear stress at the interface and the perturbation of the friction stress (the latter includes the elastodynamic bi-material effect and constitutive effects). The spectrum can feature several distinct classes of solutions.
\item The linear stability spectrum is first analyzed in the small $k H$ limit and the existence a universal instability (independent of the details of the friction law) is analytically demonstrated. The instability is shown to be related to waveguide propagating modes, which are strongly coupled to the height $H$. They are characterized by a wave-number satisfying $k H\!\sim\!{\C O}(1)$ and a growth rate that scales with $H^{-1}$.
\item As the growth rate of the waveguide-like instability vanishes in the $H\!\to\!\infty$ limit, the linear stability spectrum is analyzed also in the large $k H$ limit, where additional instabilities are sought for. Two classes of new instabilities, qualitatively different from the waveguide-like instability, are found: (i) A dynamic instability which is mediated by modes propagating at nearly the dilatational wave-speed (super-shear) in the opposite direction to the sliding motion and features a vanishingly small wave-number at threshold (ii) A dynamic instability which is mediated by modes propagating at nearly the shear wave-speed in the direction of sliding motion and features a finite wave-number at threshold.
\item In addition, a third type of instability --- which was previously discussed in the literature --- exists in the quasi-static regime.
\item In all cases, even when all wave-numbers become unstable in a certain parameter range, the frictional response is shown to be mathematically well-posed.
\item A comprehensive stability phase diagram in the large $k H$ limit is derived, presented and physically rationalized. The stability phase diagram is expressed in terms of relevant set of dimensionless parameters that quantify the competing physical effects involved.
\item A regularized, finite-time, response of the friction stress to normal stress variations is analyzed in detail and is shown to promote stability. That is, the instabilities mentioned above still exist, but their appearance is delayed and the range of unstable wave-numbers is reduced as compared to the case of an instantaneous response to normal stress variations.
\item The results may have implications for understanding the failure/rupture dynamics of a large class of bi-material frictional interfaces.
\end{itemize}

\section{Basic equations and constitutive relations}
\label{sec:math}

\begin{figure}[here]
\centering
\begin{tabular}{ccc}
\includegraphics[width=0.5\textwidth]{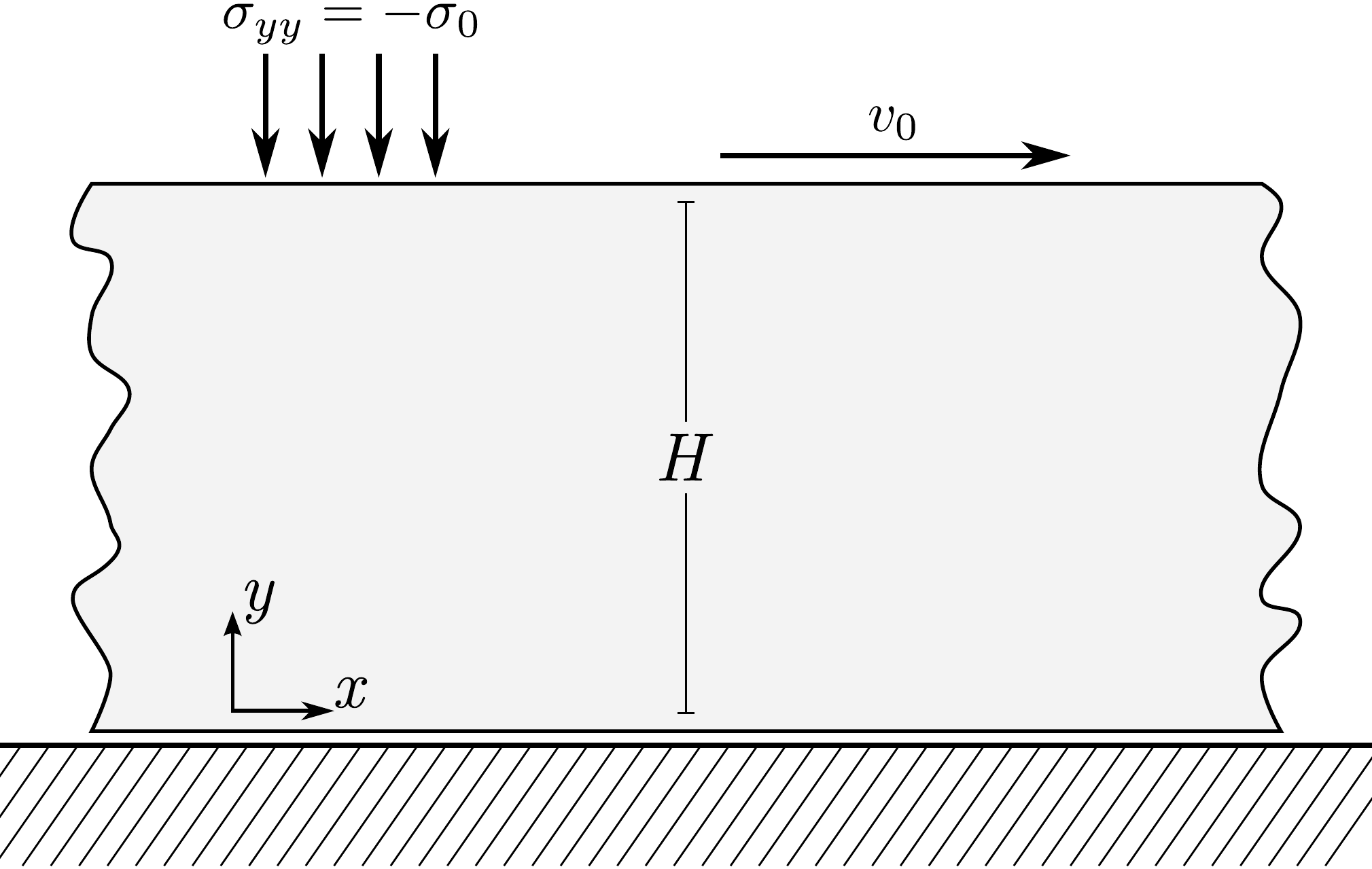}
\end{tabular}
\caption{A schematic sketch of the geometry of the system. Note that the system is regarded as infinite in the $x$-direction.}
\label{Fig1}
\end{figure}

The problem we study involves an isotropic linear elastic solid, of infinite extent in the $x$-direction and height $H$ in the $y$-direction, homogeneously sliding at a velocity $v_0$ in the $x$-direction on top of a rigid (non-deformable and stationary) half space. Two-dimensional plane-strain deformation conditions are assumed and the geometry of the problem is sketched in Fig.~\ref{Fig1}.

The deformable solid is described by the isotropic Hooke's law $\sigma_{ij}\!=\! 2\mu\varepsilon_{ij} \!+\! (K\!-\!\tfrac{2\mu}{3}) \delta_{ij} \varepsilon_{kk}$, where $\B \sigma(x,y,t)$ is Cauchy's stress tensor and the linearized strain tensor $\B \varepsilon(x,y,t)$ is derived from the displacement field $\B u(x,y,t)$ according to $\varepsilon_{ij}\!=\! \tfrac{1}{2}\left(\partial_i u_j \!+\! \partial_j u_i\right)$. $K$ and $\mu$ are the bulk and shear moduli, respectively. The bulk dynamics are determined by linear momentum balance
\begin{equation}
\label{eq:eom}
\partial_j \sigma_{ij} = \rho\,\ddot{u}_i \ ,
\end{equation}
where $\rho$ is the mass density and superimposed dots represent partial time derivatives. A uniform compressive normal stress of magnitude $\sigma_0$ is applied to the upper boundary of the sliding body
\begin{equation}
\sigma_{yy}(x,y\!=\!H,t) = -\sigma_0 \ ,
\end{equation}
where $\sigma_0\!>\!0$ and hence $\sigma_{yy}(x,y\!=\!H,t)\!<\!0$. To maintain steady sliding at a velocity $v_0$ one can either impose
\begin{equation}
\dot{u}_x(x,y\!=\!H,t)\!=\!v_0 \quad\quad\hbox{or}\quad\quad \sigma_{xy}(x,y\!=\!H,t)\!=\!\tau_0
\label{eq:2BC}
\end{equation}
such that $v_0$ emerges from the latter relation (for a given $\tau_0$, $v_0$ is determined by the friction law, see below). In the limit $H\!\to\!\infty$, these two boundary conditions are equivalent as all perturbation modes decay far from the sliding interface. This is not the case for a finite $H$, as will be discussed later.

To complete the formulation of the problem, we need to specify the boundary conditions on the sliding interface at $y\!=\!0$. First, as the lower body is assumed to be infinitely rigid, we have
\begin{equation}
u_y(x,y\!=\!0,t) = 0 \ .
\label{eq:rigidBC}
\end{equation}
Note that interfacial opening displacement, $u_y(x,y\!=\!0,t)\!>\!0$, is excluded, which is fully justified in the context of the linearized analysis to be performed below. Equation~\eqref{eq:rigidBC} is obtained in the limit of large material contrast. In the opposite limit, i.e. for interfaces separating identical materials, the boundary condition reads $\sigma_{yy}(x,y\!=\!0)\!=\!-\sigma_0$ (assuming also symmetric geometry and anti-symmetric loading). This difference in the interfacial boundary condition is responsible for all of the bi-material effects to be discussed below. The identical materials problem is qualitatively different from the bi-material problem and most of the instabilities we discuss below for the latter problem are expected not to exist for the former.

Next, we should specify a friction law which takes the form of a relation between three interfacial quantities
\begin{equation}
\sigma(x,t) \equiv -\sigma_{yy}(x,y\!=\!0,t) \ , \quad\quad \tau(x,t) \equiv \sigma_{xy}(x,y\!=\!0,t)  \ , \quad\quad v(x,t) \equiv \dot{u}_{x}(x,y\!=\!0,t) \ ,
\label{eq:frictionBC}
\end{equation}
and possibly a small set of internal state variables (note that in the general case we have $v(x,t)\!\equiv\!\dot{u}_{x}(x,y\!=\!0^+,t)\!-\!\dot{u}_{x}(x,y\!=\!0^-,t)$. Here $\dot{u}_{x}(x,y\!=\!0^-,t)\!=\!0$ due to the rigid substrate and only $y\!\ge\!0$ is of interest). As was mentioned in Sect. \ref{sec:intro}, the class of rate-and-state friction models we consider involves a single internal state variable, which we denote by $\phi(x,t)$ (the physical meaning of $\phi$ will be discussed later). Within this constitutive framework, the two dynamical interfacial fields, the friction stress $\tau(x,t)$ and $\phi(x,t)$, satisfy a coupled set of ordinary differential equations in time (no spatial derivatives are involved) of the form
\begin{equation}
\label{eq:constitutive}
\dot\tau = F(\tau,\sigma,v,\phi) \quad\quad\hbox{and}\quad\quad \dot\phi = G(\tau,\sigma,v,\phi) \ .
\end{equation}
Various explicit forms of the functions $F(\tau,\sigma,v,\phi)$ and $G(\tau,\sigma,v,\phi)$ will be discussed later (some were already alluded to in Sect. \ref{sec:intro}).

Under homogeneous steady-state conditions, when $\sigma_0$ and $v_0$ are controlled at the upper boundary of the sliding body, $\tau_0$ and $\phi_0$ are determined from Eq. (\ref{eq:constitutive}) according to
\begin{equation}
F(\tau_0,\sigma_0,v_0,\phi_0)=0 \quad\quad\hbox{and}\quad\quad G(\tau_0,\sigma_0,v_0,\phi_0) = 0 \ ,
\end{equation}
and the steady-state displacement vector field ${\B u}^{(0)}(x,y,t)$ satisfies
\begin{equation}
\label{eq:ss}
u^{(0)}_x(x,y,t) =  v_0 t + \frac{\tau_0}{\mu}\,y \qquad\hbox{and}\qquad u^{(0)}_y(x,y,t) = -\frac{\sigma_0}{K+\tfrac{4\mu}{3}}\, y \ .
\end{equation}
We assume $v_0\!>\!0$ hereafter. Our goal in the rest of this paper would be to study the linear stability of this solution, for various constitutive relations.

\section{Linear stability spectrum for standard rate-and-state friction}
\label{sec:spectrum}

We consider first the standard rate-and-state friction model in which the friction stress $\tau$ depends on the slip velocity $v$ and a state variable $\phi$~\citep{Marone1998, Baumberger2006}. The latter is of time dimensions and represents the age (or maturity) of contact asperities~\citep{Dieterich1978, Ruina1983, Baumberger2006}. It is directly related to the amount of real contact area at the interface (the ``older'' the contact, the larger the real contact area is) and hence to the strength of the interface~\citep{Dieterich1994, Nakatani2001, Nagata2008, Ben-David2010-ageing}. In static situations, when $v\!=\!0$, the age of a contact formed at time $t_0$ is simply $\phi\!=\!t-t_0$. In steady sliding at a velocity $v_0$, the lifetime of a contact of linear size $D$ is $\phi_0\!=\!D/v_0$~\citep{Dieterich1978, Teufel1978, Rice1983, Marone1998, Nakatani2001, Baumberger2006}. $D$ is a memory length scale which plays a central role in this class of models. These two limiting cases are smoothly connected by choosing the function $G(\cdot)$ in Eq.~\eqref{eq:constitutive} such that~\citep{Ruina1983, Rice1983, Marone1998, Baumberger2006}
\begin{equation}
\label{eq:rsf_phi}
\dot\phi = G(\tau,\sigma,v,\phi) = g\Big(\frac{v\,\phi}{D}\Big) \ ,
\end{equation}
with $g(1)\!=\!0$ and $g(0)\!=\!1$. Furthermore, as sliding reduces the age of the contacts, we typically expect $g'(1)\!<\!0$.

The evolution of $\tau$ is assumed to take the form
\begin{equation}
\dot\tau = F(\tau,\sigma,v,\phi) = -\frac{1}{T}\left(\tau - \sigma f(\phi,v)\right) \ .
\label{eq:gradual_sigma}
\end{equation}
For $T\!\to\!0$, which is assumed here (in Sect.~\ref{sec:modified-models} we will consider a finite $T$), we obtain
\begin{equation}
\tau = \sigma f(\phi,v) \ ,
\label{eq:rsf_stress}
\end{equation}
which completes the definition of the standard rate-and-state friction model.

Within a linear perturbation approach, the steady-state elastic fields in Eq.~\eqref{eq:ss} and the steady-state
of the state variable, $\phi_0\!=\!D/v_0$, are introduced with interfacial (i.e.~at $y\!=\!0$) perturbations proportional
to $\exp\!{[\Lambda t - i k x]}$, where $k$ is a real wave-number and $\Lambda$ is a complex growth rate.
The ultimate goal then is to find the linear stability spectrum, $\Lambda(k)$, and in particular to understand under
which conditions $\Re(\Lambda)$ changes sign. $\Re(\Lambda)\!<\!0$ corresponds to stability as perturbations decay
exponentially in time, while $\Re(\Lambda)\!>\!0$ corresponds to instability as perturbations grow exponentially.

Before we perform this analysis, we would like to gain some physical insight into the structure of the problem. For that aim, we consider Eqs. \eqref{eq:frictionBC} and \eqref{eq:rsf_stress}, and calculate the variation of the latter in the form
\begin{equation}
\delta\sigma_{xy} = \delta\tau = \sigma_0\,\delta f - f\,\delta\sigma_{yy} \ ,
\label{eq:schematic_LSA}
\end{equation}
where all quantities are evaluated at the interface, $y\!=\!0$, and the variation is taken relative to the same interfacial field (e.g. the perturbation in the slip $\delta u_x$ or slip velocity $\delta v$). When all of the terms are evaluated, the above expression becomes an implicit equation for the linear stability spectrum $\Lambda(k)$. It is obtained through a balance of three contributions. One contribution is $\delta\sigma_{xy}$, which is determined from the elastodynamic perturbation problem and is not directly coupled to the friction coefficient $f$. Another contribution is proportional to $\delta\sigma_{yy}$, which is also determined from the elastodynamic perturbation problem, but which is multiplied by the friction coefficient $f$. As this is the only place in the perturbation analysis where $f$ appears explicitly, every term that is proportional to $f$ in the expressions to follow, can be physically identified with the variation of the normal stress $\sigma_{yy}$ (and hence with the bi-material effect). Finally, the remaining contribution is determined by the variation of the friction coefficient $\delta f(\phi,v)$, which is affected both by the variation of the slip velocity $v$ and of the state variable $\phi$, and is multiplied by the applied normal stress $\sigma_0$. Next, we aim at calculating each of these contributions, which are being intentionally presented in some detail.

\subsection{Perturbation of the elastodynamic fields}

We seek a solution of the linear momentum balance Eqs. \eqref{eq:eom}, coupled to Hooke's law, which is proportional to $\exp\!{[\Lambda t - ikx]}$.
Consequently, we derive the general solution $\delta{\B u}(x,y,t)$ of the problem as a superposition of shear-like and dilatational-like modes (which lead to an interfacial perturbation proportional to $\exp\!{[\Lambda t - ikx]}$) in the form
\begin{equation}
\begin{pmatrix}
 \delta u_x(x,y,t)\\ \delta u_y(x,y,t)
 \end{pmatrix}
 =
 \begin{pmatrix}
 -k_s & k & k_s & k \\
 ik  & - ik_d & ik & i k_d
 \end{pmatrix}
  \begin{pmatrix}
   A_1 \exp[-k_s y] \\
   A_2 \exp[-k_d y] \\
   \hspace{-0.3cm}A_3 \exp[k_s y] \\
   \hspace{-0.3cm}A_4 \exp[k_d y]
  \end{pmatrix}
  \exp\!{[\Lambda t - ikx]} \ ,
\label{eq:elastic fields}
\end{equation}
where $\{A_i\}_{i=1-4}$ are yet undetermined amplitudes and we defined $k_s(\Lambda,k)\!\equiv\!\sqrt{\Lambda^2/c_s^2 + k^2}$ and $k_d(\Lambda,k)\!\equiv\!\sqrt{\Lambda^2/c_d^2 + k^2}$.
The shear and dilatational wave-speeds are $c_s\!=\!\sqrt{\mu/\rho}$ and $c_d\!=\!\sqrt{\tfrac{3K + 4\mu}{3\rho}}$, respectively.
Note that $k_{s,d}$ are in general complex (since $\Lambda$ is complex) and that since we adopt the convention that the branch-cut of the complex square root function
lies along the negative real axis, we have $\Re(k_{s,d})\!\ge\!0$.

To proceed, we need to impose physically relevant boundary conditions. Recall that at $y\!=\!0$ we have $\delta u_y(x,y\!=\!0,t)\!=\!0$ due to the presence of an infinitely rigid substrate. We then focus on the case in which the velocity is controlled at $y\!=\!H$, i.e.~$\dot{u}_x(x,y\!=\!H,t)\!=\!v_0$, cf. Eq. \eqref{eq:2BC}. Consequently, as both the normal stress $\sigma_{yy}$ and tangential displacement $u_x$ are controlled at $y\!=\!H$,
the perturbation satisfies $\delta\sigma_{yy}(x,y\!=\!H,t)\!=\!\delta u_x(x,y\!=\!H,t)\!=\!0$.
Imposing these three boundary conditions on the general solution in Eq. \eqref{eq:elastic fields} allows us to eliminate the four amplitudes $\{A_i\}_{i=1-4}$
in favor of a single undetermined amplitude, and express $\delta\sigma_{xy}$ and $\delta\sigma_{yy}$ at $y\!=\!0$ in terms of $\delta u_x$.
Using Hooke's law to transform the displacement field into the interfacial stress components, we obtain
\begin{equation}
\label{eq:pert_elastodynamics}
\delta\sigma_{xy}=-\mu\,k_d(\Lambda,k)\,G_1(\Lambda,k, H)\,\delta u_x \quad\quad\hbox{and}\quad\quad \delta\sigma_{yy} =i\,\mu\,k\,G_2(\Lambda,k, H)\,\delta u_x \ ,
\end{equation}
with
\begin{equation}
\label{eq:pert_elastodynamics_auxiliary}
G_1(\Lambda, k, H)\equiv\frac{\coth(Hk_d)\Lambda ^2/c_s^2 }{k_d k_s \coth(Hk_d)\tanh(Hk_s)-k^2}\qquad\hbox{and}\qquad G_2(\Lambda, k, H)\equiv2- \frac{G_1(\Lambda, k, H)}{\coth(Hk_d)} \ ,
\end{equation}
and recall that $k_{s,d}$ are both functions of $\Lambda$ and $k$. This completes the elastodynamic calculation of $\delta\sigma_{xy}$ and $\delta\sigma_{yy}$ at the interface.

\subsection{Perturbation of the friction law}

In the next step, we consider the perturbation of the friction law. The calculation is straightforward, yet for completeness and full transparency we explicitly present it here. The variation of $f(\phi,v)$ with respect to slip velocity perturbations takes the form
\begin{equation}
\delta f = \left(\pa_v\!f + \pa_{\phi} f \frac{\delta\phi}{\delta v} \right) \delta v \ ,
\label{eq:delta_f1}
\end{equation}
where all of the derivatives are evaluated at the steady-state values corresponding to $v_0$. As the perturbation in $\phi$ takes the
 form $\delta\phi\!\sim\!\exp\!{[\Lambda t - ikx]}$, we can use Eq. \eqref{eq:rsf_phi} to obtain
\begin{equation}
\frac{\delta\phi}{\delta v} = \frac{g'(1)}{\displaystyle v_0 \left(\Lambda -g'(1) \frac{v_0}{D} \right)} = -\frac{|g'(1)|}{\displaystyle v_0 \left(\Lambda +|g'(1)| \frac{v_0}{D} \right)} \ ,
\label{eq:delta_phi}
\end{equation}
where in the latter we used $g'(1)\!=\!-|g'(1)|$ because $g'(1)\!<\!0$.
Finally, since $d_v \phi_0\!=\!-D/v_0^2$, we can relate $\pa_{\phi} f$ to $d_v\!f$ according to
\begin{equation}
d_v\!f = \pa_v\!f - \frac{D}{v_0^2}\pa_{\phi} f \ .
\label{eq:df}
\end{equation}
Substituting Eqs.~\eqref{eq:delta_phi}-\eqref{eq:df} into Eq. \eqref{eq:delta_f1} and using $\delta v\!=\!\Lambda \delta u_x$, we obtain the final expression for the perturbation of $f(\phi,v)$
\begin{equation}
\delta f = \frac{\Lambda}{\displaystyle \Lambda + \frac{|g'(1)|\,v_0}{D}}\left(\Lambda\,\pa_v\!f + \frac{|g'(1)|\,v_0}{D}\,d_v\!f\right) \delta u_x \ .
\label{eq:delta_f2}
\end{equation}

\subsection{The linear stability spectrum and dimensionless parameters}

Now that we have calculated the perturbations of $\delta\sigma_{xy}$, $\delta\sigma_{yy}$ and $\delta f$ in terms of $\delta u_x$, we are ready to derive the linear stability spectrum. For that aim, we substitute Eqs. \eqref{eq:pert_elastodynamics} and \eqref{eq:delta_f2} into Eq. \eqref{eq:schematic_LSA} and eliminate $\delta u_x$ to obtain
\begin{align}
\hspace{-2cm}S(\Lambda, k, H) \!\equiv\! \mu\,k_d(\Lambda,k)\,G_1(\Lambda,k, H) \!-\! i\,\mu\,k\,f\,G_2(\Lambda,k, H)  \!+\! \frac{\sigma_0\,\Lambda}{\Lambda + \frac{|g'(1)|v_0}{D}}
\left( \Lambda \pa_v\!f + \frac{|g'(1)|v_0}{D} d_v\!f \right) \!=\! 0 \ .
\hspace{-2cm}\nonumber\\
\label{eq:spectrum_dim}
\end{align}
This is an implicit expression for $\Lambda(k)$, which is in general a multi-valued function with various branches. Analyzing Eq.~\eqref{eq:spectrum_dim} is the major goal of this paper.

As a first step, we briefly discuss the symmetry properties of Eq.~\eqref{eq:spectrum_dim}.
The latter satisfies $S(\overline{\Lambda}, -k, H)\!=\!\overline{S(\Lambda, k, H)}$, where a bar denotes complex conjugation. This implies that for each mode with $k\!<\!0$, there exists another mode with $k\!>\!0$, which has the same growth rate $\Re(\Lambda)$ and an opposite sign frequency $-\Im(\Lambda)$. Consequently, we have $\Lambda(-k)\!=\!\overline{\Lambda(k)}$. These two modes have the same phase velocity $\Im(\Lambda)/k$ (because both the numerator and denominator change sign when $k\!\to\!-k$) and they propagate in the same direction. These symmetry properties allow us to assume hereafter $k\!>\!0$ without loss of generality.

We would now like to discuss the set of dimensionless parameters that control the linear stability problem. In addition to $f$ itself, which is obviously a relevant dimensionless quantity, we define the following three quantities
\begin{equation}
\Delta \equiv \frac{d_v\!f}{\pa_v\!f}\ , \quad\quad \beta \equiv \frac{c_s}{c_d}\ , \quad\quad \gamma \equiv \frac{\mu}{c_s\,\sigma_0\,\pa_v\!f} \ ,
\label{eq:def_dimless}
\end{equation}
which involve material/interfacial parameters, interfacial constitutive functions that may depend on the sliding velocity (i.e.~$d_v\!f$ and $\pa_v\!f$), and the applied normal stress $\sigma_0$.  The first quantity, $\Delta$, is the ratio between the steady-state and the instantaneous variation of $f$ with $v$ (``instantaneous'' here means that the slip velocity variation takes place on a time scale over which the state variable $\phi$ does not change appreciably). As such, $\Delta$ is a property of the frictional interface. As mentioned above, frictional interfaces generally exhibit a positive instantaneous response to slip velocity changes (a positive ``direct effect''), $\pa_v\!f\!>\!0$,
which is assumed here. Consequently, the sign of $\Delta$ is determined by the sign of $d_v\!f$, that is, by whether friction is steady-state velocity-weakening or velocity-strengthening in a certain range of sliding velocities.
While steady-state velocity-weakening is prevalent at small sliding velocities and is very important for rapid slip nucleation \citep{Rice1983, Dieterich1992, Ben-Zion2008}, some frictional interfaces are intrinsically steady-state velocity-strengthening~\citep{Noda2009, Ikari2009, Ikari2013}. Moreover, it has been recently argued~\citep{Bar-Sinai2014jgr} that dry frictional interfaces generically exhibit a crossover from steady-state velocity-weakening, $d_v\!f\!<\!0$, to velocity-strengthening, $d_v\!f\!>\!0$, with increasing $v$ beyond a local minimum. This claim has been supported by a rather extensive set of experimental data for a range of materials~\citep{Bar-Sinai2014jgr}.

From this perspective, it would be instructive to write $\Delta$ using Eq.~\eqref{eq:df} as
\begin{equation}
\Delta = 1 - \frac{D\,\pa_{\phi} f}{v_0^2\, \pa_v\!f} = 1 - \frac{\pa_{_{\log{\!\phi}}} f}{\pa_{_{\log{\!v}}} f} \ ,
\label{eq:Delta}
\end{equation}
where generically $\pa_{\phi} f\!>\!0$ (i.e.~frictional interfaces are stronger the older -- the more mature -- the contact is) and recall that all derivatives are evaluated at steady-state corresponding to a sliding velocity $v_0$ (i.e.~$v\!=\!v_0$ and $\phi_0\!=\!D/v_0$). Hence, a crossover from $\Delta\!<\!0$ at relatively small $v_0$'s to $\Delta\!>\!0$ at higher $v_0$'s implies a crossover from $\pa_{_{\log{\!\phi}}} f\!>\!\pa_{_{\log{\!v}}} f$ to $\pa_{_{\log{\!\phi}}} f\!<\!\pa_{_{\log{\!v}}} f$ with increasing $v_0$. While our analysis can be in principle applied to any $\Delta$, the most interesting physical regime corresponds to $\Delta\!>\!0$, i.e.~to sliding on a steady-state velocity-strengthening friction branch, where friction appears to be stabilizing and the destabilization emerges from elastodynamic bi-material and finite size effects (see below).
For steady-state velocity-weakening friction, which has been studied quite extensively in the past~\citep{Dieterich1978, Rice1983, Marone1998},
we have $\Delta\!<\!0$ and some unstable modes {\em always} exist.
This instability is of a different physical origin compared to the instabilities to be discussed below.
Physical considerations~\citep{Bar-Sinai2014jgr} indicate that increasing the steady-state sliding velocity $v_0$ along
the velocity-strengthening branch is accompanied by either a decrease in $\pa_{_{\log{\!\phi}}} f$
or an increase in $\pa_{_{\log{\!v}}} f$, or both. In the limiting case, $\pa_{_{\log{\!\phi}}} f\!\ll\!\pa_{_{\log{\!v}}} f$,
we have $\Delta\!\to\!1$. Consequently, on the steady-state velocity-strengthening branch we have $0\!\le\!\Delta\!\le\!1$.

The second quantity, $\beta$, is the ratio between the shear wave-speed $c_s$ and the dilatational wave-speed $c_d$ and hence is a purely linear elastic (bulk) quantity. $\beta$ is only a function of Poisson's ratio $\nu$ (in terms of the shear and bulk moduli, we have $\nu\!=\!\tfrac{3K-2\mu}{6K+2\mu}$), which for plane-strain conditions takes the form
\begin{equation}
\beta = \sqrt{\frac{1-2\nu}{2(1-\nu)}} \ .
\label{eq:beta}
\end{equation}
For ordinary materials we have $0\!\le\!\nu\!\le\!\tfrac{1}{2}$ (recall that thermodynamics imposes a broader constraint, $-1\!\le\!\nu\!\le\!\tfrac{1}{2}$, but negative Poisson's ratio materials are excluded from the discussion here), which translates into $0\!\le\!\beta\!\le\!\tfrac{1}{\sqrt{2}}$.

The third quantity, $\gamma$, is the ratio between the elastodynamic quantity $\mu/c_s$ --- proportional to the so-called radiation damping factor for sliding~\citep{Rice1993, Rice2001, Crupi2013} --- and the instantaneous response of the frictional stress to variations in the sliding velocity, $\pa_v\tau\!=\!\sigma_0\pa_v\!f$. The latter is a product of the externally applied normal stress $\sigma_0$ and $\pa_v\!f$. As such, $\gamma$ quantifies the relative importance of elastodynamics, the applied normal stress and the direct effect (an intrinsic interfacial property). It has been shown that many frictional interfaces are characterized by an instantaneous linear dependence of $f$ on $\log{v}$ over some range of slip velocities due to thermally-activated rheology. In this case, $\pa_{_{\log{\!v}}} f$ is a positive constant and $\pa_v\!f\!=\!\tfrac{\pa_{_{\log{\!v}}} f}{v_0}$ is a decreasing function of $v_0$. Therefore, as $v_0$ increases (for a fixed $\sigma_0$), elastodynamics becomes more important and $\gamma$ increases. Finally, note that as $\sigma_0\!>\!0$ and $\pa_v\!f\!>\!0$, we have $\gamma\!>\!0$.

In terms of the four independent dimensionless parameters $\Delta$, $\beta$, $\gamma$ and $f$, the linear stability spectrum in Eq.~\eqref{eq:spectrum_dim} can be rewritten as
\begin{eqnarray}
S(\Lambda, k, H) =\mu\,k_d(\Lambda,k)\,G_1(\Lambda,k, H) - i\,\mu\,k\,f\,G_2(\Lambda,k, H)  + \mu \frac{\Lambda}{c_s} \frac{\lambda(\Lambda)+\Delta}{\gamma\left(\lambda(\Lambda)+1\right)} = 0 \ ,
\label{eq:Hspectrum_dimensionless}
\end{eqnarray}
where $\lambda(\Lambda)\!\equiv\!\tfrac{\Lambda D}{|g'(1)|v_0}$. The shear modulus $\mu$ clearly drops as
a common factor and then the only dimensional quantities left are $k$ and $\Lambda/c_s$,
which obviously form a dimensionless combination.

The main goal of most of the remaining parts of this paper is to find the solutions (i.e. roots) of Eq.~\eqref{eq:Hspectrum_dimensionless}, especially those with $\Re(\Lambda)\!>\!0$. Equation~\eqref{eq:Hspectrum_dimensionless} is a complex equation (both in the complex-variable sense and in the literal sense) which includes branch-cuts in the complex-plane and is expected to have several solutions $\Lambda(k)$. In looking for these solutions, one can follow various strategies. One strategy would be to numerically search for these solutions. We do not adopt this strategy in this paper. Rather, we will show that solutions of Eq.~\eqref{eq:Hspectrum_dimensionless} are physically related to various elastodynamic solutions and that this insight can be used to obtain a variety of analytic results (which are then supported numerically). Below we treat separately the small and large $k H$ limits of Eq.~\eqref{eq:Hspectrum_dimensionless}.

\section{Analysis of the spectrum in the small $k H$ limit}
\label{sec:finiteH}

Our first goal is to analyze the linear stability spectrum in Eq.~\eqref{eq:Hspectrum_dimensionless} in the small $k H$ limit, where $k H\!\sim\!{\C O}(1)$ or smaller.
In this range of wave-numbers $k$, perturbations are strongly coupled to the finite boundary at $y\!=\!H$.
The most notable physical implication of this is that the elastodynamic solutions in the bulk are non-decaying in the $y$-direction.
Consequently, we will look for solutions that are related to waveguide-like solutions featuring a propagative wave nature in the $x$-direction and
a standing wave nature in the $y$-direction.

In general, a 2D wave equation would give rise to a dispersion relation of the form $\Lambda^2\!=\!-c^2(k_x^2 + k_y^2)$ ($c$ is some wave-speed).
In the presence of a finite boundary at $y\!=\!H$, $k_x\!\equiv\!k$ is continuous and $k_y$ is quantized according to $k_y\!=\!m\,\pi/H$,
with $m$ being a set of integers/half-integers which are determined from the boundary conditions.
This quantization implies that in the long wavelength limit $k\!\to\!0$, the dispersion relation results in a cutoff frequency
$\Lambda\!=\!\pm i c \,m\,\pi/H$. This property will be shown below to have significant implications on the stability problem.

Based on the idea that waveguide-like solutions might be important for the stability problem, we seek now solutions
to Eq.~\eqref{eq:Hspectrum_dimensionless} in the limit $k\!\to\!0$. For that aim, we look first for solutions corresponding to
$\delta\sigma_{xy}(\Lambda, k)\!=\!0$, which defines the relevant waveguide dispersion relation $\Lambda_{wg}(k)$.
Following Eqs.~\eqref{eq:pert_elastodynamics}-\eqref{eq:pert_elastodynamics_auxiliary}, solutions of interest correspond to $\coth(Hk_d)\!=\!0$, i.e.
\begin{equation}
H k_d(\Lambda_{wg},k)=H \sqrt{\frac{\Lambda_{wg}^2}{c_d^2} + k^2}=\pm \frac{i(2n+1)\pi}{2}\qquad\Longrightarrow\qquad
\Lambda_0 \equiv \Lambda_{wg}(k\!\to\!0) = \pm \frac{i(2n+1)\pi c_d}{2H} \ ,
\label{eq:waveguide_cutoff}
\end{equation}
where $\Lambda_0$ is the waveguide cutoff frequency.

Our strategy will be to expand $S(\Lambda, k, H)$ of Eq.~\eqref{eq:Hspectrum_dimensionless},
as a function of the two variables $k$ and $\Lambda$, to leading order around $(\Lambda\!=\!\Lambda_0, k\!=\!0)$.
That is, we expand $S(\Lambda, k, H)$ to linear order in $k$ and in $\delta\Lambda$, $\Lambda\!\simeq\!\Lambda_0+\delta\Lambda+{\C O}(\delta\Lambda^2)$,
and then set $S(\Lambda, k, H)\!=\!0$ to obtain $\delta\Lambda(k)$. This procedure is expected to yield a solution that satisfies $\delta\Lambda(k)\!\sim\!k$. In principle, if indeed a solution that satisfies $\Re[\delta\Lambda(k)]\!\sim\!k$ exists (i.e.~$k\!=\!0$ is a regular point where a linear expansion exists), then irrespective of the sign of the proportionality coefficient we have $\Re[\delta\Lambda(k)]\!>\!0$ for either $k\!>\!0$ or $k\!<\!0$, implying an instability. In fact, a similar conclusion can be reached even if the discussion is restricted to $k\!>\!0$. In this case, if $\Lambda(k)\!\simeq\!\Lambda_0\!+\!\delta\Lambda(k)$ ---
with a pure imaginary $\Lambda_0$ and a complex $\delta\Lambda(k)\!\sim\!k$ --- is a solution of $S(\Lambda(k), k, H)\!=\!0$ then the symmetry property
$S(\overline{\Lambda}, -k, H)\!=\!\overline{S(\Lambda, k, H)}$ implies that $S(\bar{\Lambda}_0+\Re[\delta\Lambda(k)]-i \Im[\delta\Lambda(k)], -k, H)=0$.
Therefore,
\begin{equation}
\Lambda(k)\simeq\bar{\Lambda}_0+\Re[\delta\Lambda(-k)]-i \Im[\delta\Lambda(-k)]=
\bar{\Lambda}_0-\Re[\delta\Lambda(k)]+i \Im[\delta\Lambda(k)]
\end{equation}
is also a solution of $S(\Lambda(k), k, H)\!=\!0$.

We thus conclude, based on the last argument, that if a solution with a complex
$\delta\Lambda(k)\!\sim\!k$ near $\Lambda_0$ (purely imaginary) exists, then actually there exist {\em two} solutions with the {\em same} $\Im[\delta\Lambda(k)]$
and $\Re[\delta\Lambda(k)]$ of {\em opposite} signs. Both solutions propagate with the same group velocity $d\Im[\delta\Lambda]/dk$,
but one is stable ($\Re[\delta\Lambda(k)]\!<\!0$) and the other is unstable ($\Re[\delta\Lambda(k)]\!>\!0$). This leads to
the quite remarkable conclusion that steady-state sliding along (strongly) bi-material frictional interfaces
in this broad class of constitutive models is {\em universally} unstable.

To fully establish this important result, we derive it by an explicit calculation. To that aim, as explained above, we need to expand $S(\Lambda, k, H)$ of Eq.~\eqref{eq:Hspectrum_dimensionless} to leading order around $(\Lambda\!=\!\Lambda_0, k\!=\!0)$. This is done in detail in Appendix~\ref{appendix_derivation}, but the essence is given here. The contribution related to $\delta\sigma_{yy}$ in Eq.~\eqref{eq:Hspectrum_dimensionless} is proportional to both $G_2(\Lambda,k, H)$ and $k$; the former diverges in the limit $(\Lambda, k)\!\to\!(\Lambda_0,0)$, while the latter clearly vanishes. In total, the term proportional to $\delta\sigma_{yy}$ approaches the finite limit
\begin{align}
i\,k\,f\,G_2(\Lambda,k, H) & \simeq \pm \frac{f\,c_s\,k}{\beta^2 H \tan(H|\Lambda_0|/c_s) \delta\Lambda} \ ,
\end{align}
where the $\pm$ corresponds to $\Lambda_0\!=\!\pm\, i |\Lambda_0|$. This leads to the non-trivial situation in which the leading order contribution involves the ratio of $\delta\Lambda$ and $k$.

Consequently, the contributions related to $\delta\sigma_{xy}$ and $\delta f$ can be taken, albeit with some care, to zeroth order, yielding (see Appendix~\ref{appendix_derivation} for details)
\begin{align}
&k_d(\Lambda,k)\,G_1(\Lambda,k, H)  \simeq \frac{|\Lambda_0|}{c_s\,\tan(H|\Lambda_0|/c_s)} \ ,\\
&\sigma_0\delta f\simeq \frac{\Lambda_0}{c_s}\frac{\lambda_0+\Delta}{\gamma\big(\lambda_0+1\big)} = \frac{|\Lambda_0|}{c_s \gamma}\left(\frac{-|\lambda_0|(1-\Delta)\pm i (\Delta +|\lambda_0|^2)}{1+|\lambda_0|^2}\right) \ ,
\label{eq:expansion_friction}
\end{align}
where $\lambda_0\!\equiv\!\tfrac{D\Lambda_0}{|g'(1)|v_0}$. Collecting all three contributions we end up with
\begin{equation}
S(\Lambda, k, H) \simeq \frac{|\Lambda_0|}{c_s\tan(H|\Lambda_0|/c_s)}
 \pm \frac{f\,c_s\,k}{\beta^2 H \tan(H|\Lambda_0|/c_s) \delta\Lambda}
 +\frac{\Lambda_0}{c_s}\frac{\lambda_0+\Delta}{\gamma\big(\lambda_0+1\big)} = 0 \ .
\label{eq:Hspectrum_expansion}
\end{equation}

Equation~\eqref{eq:Hspectrum_expansion} clearly establishes a linear relation between $\delta\Lambda$ and $k$. Extracting $\Lambda(k)\!=\!\Lambda_0 + \delta\Lambda(k)$ by solving Eq.~\eqref{eq:Hspectrum_expansion}, we obtain
\begin{equation}
\Re(\Lambda)  \simeq \mp \frac{\displaystyle\frac{f\,c_s\,k}{\beta^2 H \tan(H|\Lambda_0|/c_s)}
\left[\frac{|\Lambda_0|}{c_s\tan(H|\Lambda_0|/c_s)}-\frac{|\Lambda_0|}{c_s \gamma}\frac{|\lambda_0|(1-\Delta)}
{1+|\lambda_0|^2}\right]}{\displaystyle\left(\frac{|\Lambda_0|}{c_s\tan(H|\Lambda_0|/c_s)}-\frac{|\Lambda_0|}{c_s \gamma}\frac{|\lambda_0|(1-\Delta)}
{1+|\lambda_0|^2} \right)^2+\left( \frac{|\Lambda_0|}{c_s \gamma}\frac{\Delta+|\lambda_0|^2}
{1+|\lambda_0|^2}\right)^2} + {\C O}(k^2) \ ,
\label{eq:Hspectrum}
\end{equation}
where $\Im[\delta\Lambda]$ can be easily obtained as well and is independent of the sign of $\Im[\Lambda_0]$.
Equation~\eqref{eq:Hspectrum} has precisely the predicted structure, i.e.~there are two solutions for $\Lambda$
in the small $k\!>\!0$ limit, whose real parts have opposite signs. Consequently, a solution with $\Re[\Lambda]\!>\!0$ always exists,
i.e.~the system is universally unstable.

\begin{figure}[here]
\centering
\begin{tabular}{ccc}
\includegraphics[width=0.51\textwidth]{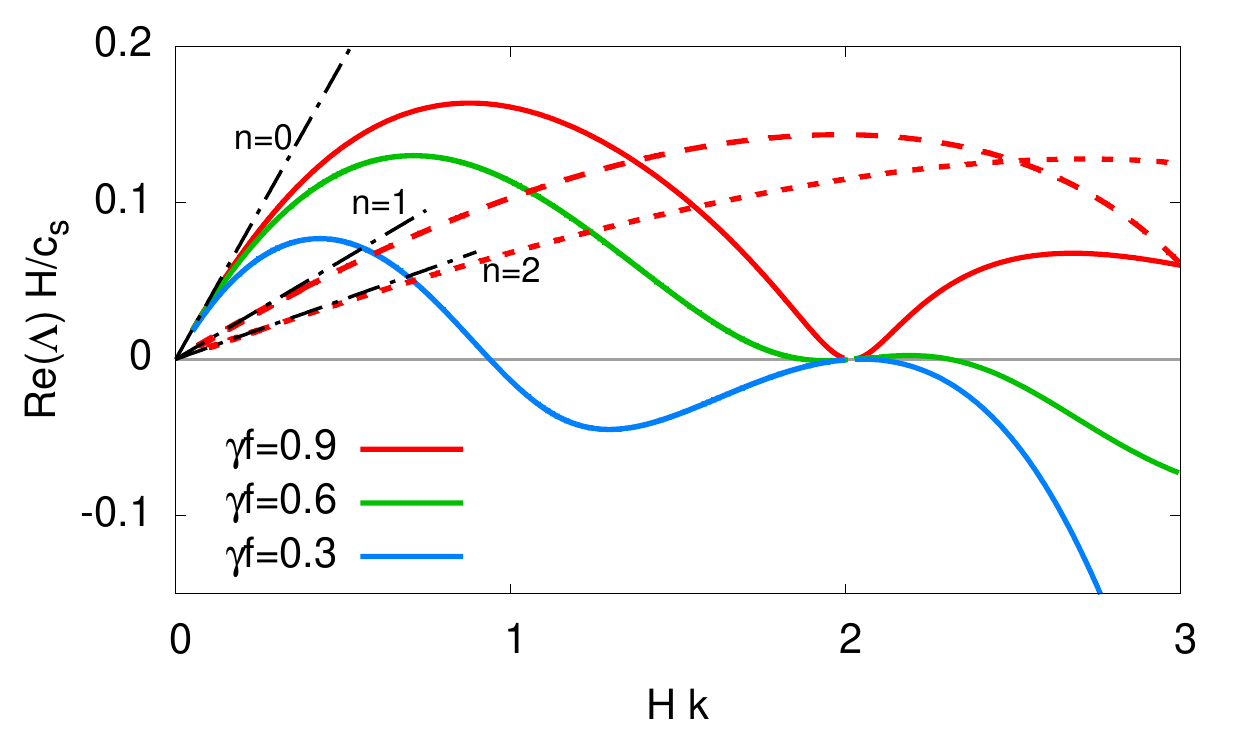}
\includegraphics[width=0.45\textwidth]{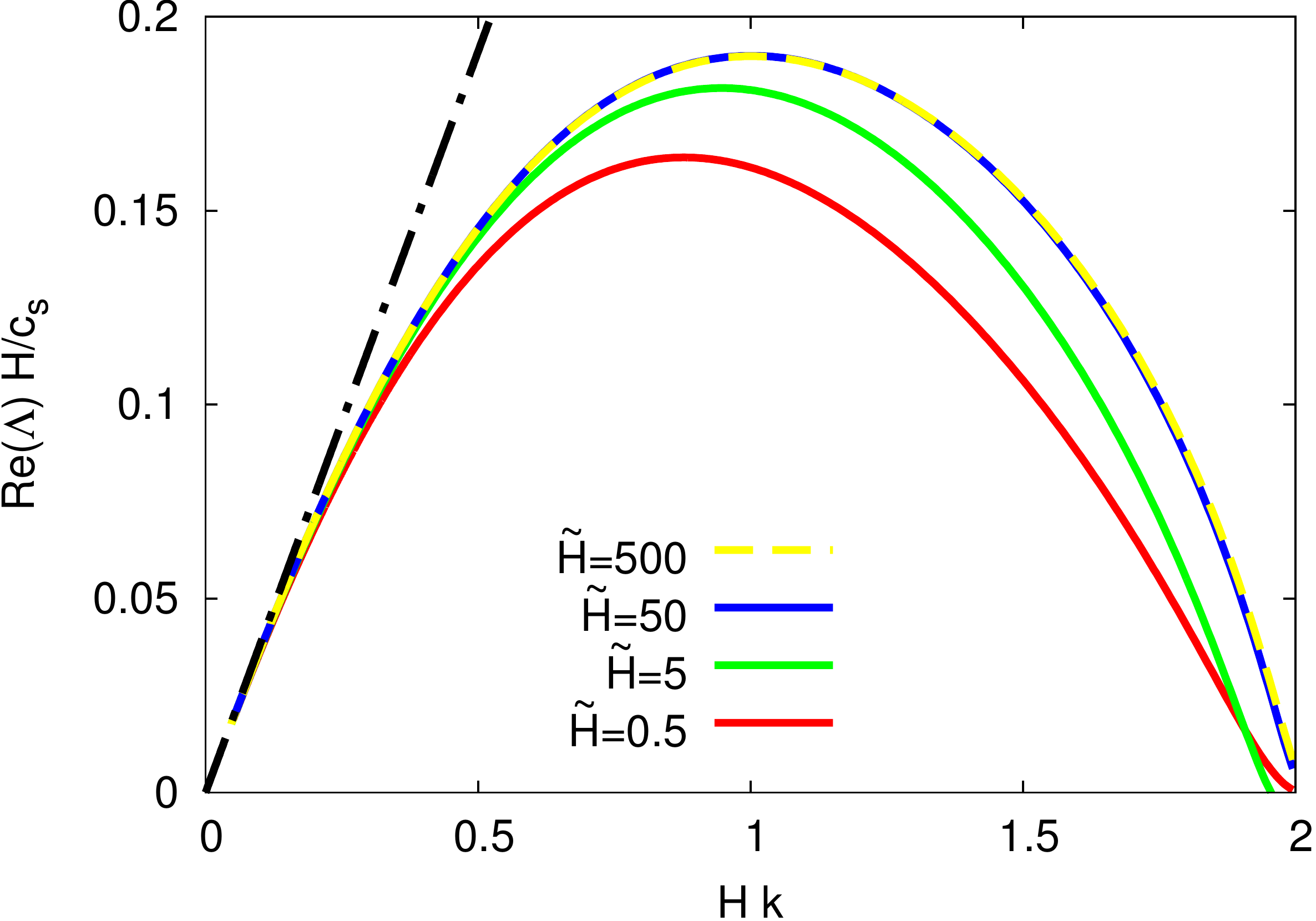}
\end{tabular}
\caption{(left) The growth rate $\Re[\Lambda]$ (in units of $c_s/H$) vs. $H k$, for small $H k$, for various values of $\gamma f$ (solid lines) and quantization numbers $n$ (dashed and dotted lines). $\gamma f$ was varied by varying $\gamma$ and we used $\tilde{H}\!=\!0.5$, where $\tilde H$ is $H$ measured in units of $c_s D/|g'(1)|v_0$, which is the product of the basic time scale of the system, $D/|g'(1)|v_0$, and the velocity $c_s$. The black broken lines show the asymptotic behavior at $k\!\to\!0$, as predicted by Eq.~\eqref{eq:Hspectrum}. (right) The same as the left panel, but for varying $H$ (and $n\!=\!0$). Note the saturation for large $H$, as predicted analytically. Unless specified otherwise, all calculations, here and in what follows, were done with the generic parameter values $f\!=\!0.3,\ \gamma\!=\!3,\ \beta\!=\!0.5$ and $\Delta\!=\!0.7$.}
\label{fig:finiteH}
\end{figure}

While Eq.~\eqref{eq:Hspectrum} can be somewhat simplified, it is retained in this form so that the physical origin of the various terms will remain transparent. Note that in the particular case of $\beta\!=\!1/2$ (which will be used in some of the numerical calculations below), a significant simplification is obtained, leading to an unstable branch $\Re(\Lambda)\!\simeq\! 4\,f\,k\,c_s/[(2n+1)\pi]+ {\C O}(k^2)$ (some care should be taken when obtaining this result as a naive substitution of $\beta\!=\!1/2$ in Eq.~\eqref{eq:Hspectrum} results in some apparently divergent contributions). The analytic result for the small $k$ behavior of the growth rate $\Re[\Lambda]$ presented in Eq.~\eqref{eq:Hspectrum}, which is one of the main results of this paper, is verified  in Fig.~\ref{fig:finiteH} for the few first $n$'s by a direct numerical solution of the linear stability spectrum in Eq.~\eqref{eq:Hspectrum_dimensionless}. The numerical solution of the spectrum shows that the most unstable mode satisfies $k H\!\sim\!{\C O}(1)$, where $\Re[\Lambda]$ attains its first maximum (corresponding to the $n\!=\!0$ solution). This can be analytically obtained by calculating the ${\C O}(k^2)$ correction
to Eq.~\eqref{eq:Hspectrum}, though the calculation is lengthy.

We thus conclude that the finite size $H$ of the sliding system has significant implications for its
stability, in particular it implies the existence of an instability with a wavelength determined by $H$.
This instability should be relevant to a broad range of systems, for example an elastic brake pad sliding over a much stiffer substrate, for which recent numerical results demonstrated dominant instability modes directly related to the intrinsic vibrational modes of the pad~\citep{Behrendt2011, Meziane2007}. The universal existence of this finite $H$ instability does not immediately mean that it will be indeed observed since other instabilities, which do not necessarily satisfy $k H\!\sim\!{\C O}(1)$, might exist and feature a larger growth rate (when several instabilities coexist, the one with the largest growth rate will be the dominant one).

To further clarify this point, we consider large, but finite, $H$ in Eq.~\eqref{eq:Hspectrum}. In this limit, by counting powers of $H$ and substituting
$k\!\sim\!H^{-1}$ for the fastest growing mode, we obtain for the latter $\Re[\Lambda]\!\propto\!f c_s/H$. This scaling is verified numerically in Fig.~\ref{fig:finiteH} (right panel). This result shows that this instability depends on the presence of friction, but not very much on the details of the friction law (e.g. the length scale $D$ does not play a dominant role), and that the growth rate of the instability decreases with increasing $H$. This raises the question of whether in the large $H$ limit there exist other instabilities with a larger growth rates, which will be addressed in the rest of the paper.

It is important to note that while we highlight the role of the finite size $H$ in relation to the universal instability encapsulated in the growth rate in Eq.~\eqref{eq:Hspectrum}, we should stress that the bi-material effect remains an essential physical ingredient driving this instability. This is evident from the observation that $\Re[\Lambda]$ is proportional to $f$ in Eq.~\eqref{eq:Hspectrum}, where --- as explained in Sect.~\ref{sec:spectrum} --- the latter is a clear signature of the variation of the normal stress with slip, which is associated with the bi-material effect. We thus expect that this universal instability does not exist for frictional interfaces separating identical materials. The effect of finite material contrast should be assessed in the future.

Before we discuss the large $H$ limit, $k H\!\to\!\infty$, we would like to note another interesting implication of the finite system size $H$. For infinite systems, $H\!\to\!\infty$, there is an equivalence between velocity-controlled and stress-controlled external boundary conditions (cf. Eq.~\eqref{eq:2BC})
because perturbations decay exponentially away from the interface in the $y$-direction. This equivalence breaks down for a finite $H$.
The analysis above focussed on velocity-controlled boundary conditions. In Appendix~\ref{appendix_BC}, we consider also stress-controlled boundary conditions and explicitly demonstrate the inequivalence of the two types of boundary conditions for finite size systems. The differences, though, are quantitative in nature and the generic instabilities discussed above remain qualitatively unchanged.

Finally, we would like to note that there can be solutions to Eq.~\eqref{eq:Hspectrum_dimensionless} in the small $k H$ limit other than Eq.~\eqref{eq:Hspectrum}. We are not looking for them here because the result in Eq.~\eqref{eq:Hspectrum} already shows that the system is always unstable. In addition, we expect the decay of the growth rate $\Re[\Lambda]$ with $H$ to be a generic property of unstable solutions of Eq.~\eqref{eq:Hspectrum_dimensionless} in the small $k H$ limit. Consequently, we focus next on the large $k H$ limit, looking for qualitatively different unstable solutions.

\section{Analysis of the spectrum in the $k H\!\to\!\infty$ limit}
\label{sec:infiniteH}

After analyzing the linear stability spectrum of Eq.~\eqref{eq:Hspectrum_dimensionless} in the small $k H$ limit, our goal now is to provide a thorough analysis of the opposite limit,
$k H\!\to\!\infty$. The length $H$ enters the problem through the elasticity relations in Eq.~\eqref{eq:pert_elastodynamics} and more precisely through the functions $G_{1,2}$ in
Eq.~\eqref{eq:pert_elastodynamics_auxiliary}. Taking the $k H\!\to\!\infty$ limit in Eq.~\eqref{eq:pert_elastodynamics_auxiliary}, which amounts to taking the arguments of $\coth(\cdot)$
and $\tanh(\cdot)$ to be arbitrarily large, we obtain
\begin{eqnarray}
\label{eq:g12}
G_1(\Lambda,k,H\!\to\!\infty) \qquad&\to&\qquad g_1(\Lambda,k)=\frac{\Lambda^2}{c_s^2 \left[k_s(\Lambda,k)\,k_d(\Lambda,k)-k^2\right]} \ ,\nonumber\\
G_2(\Lambda,k,H\!\to\!\infty) \qquad&\to&\qquad g_2(\Lambda,k)= 2-g_1(\Lambda,k) \ ,
\end{eqnarray}
which should be used in Eq.~\eqref{eq:Hspectrum_dimensionless}. The friction part is of course independent of $H$.

To further simplify the analysis of the spectrum in this limit, we define an auxiliary (and dimensionless) complex variable $z$ that relates the spatial and temporal
properties of perturbations according to
\begin{equation}
z \equiv -\frac{\Lambda}{i k\,c_s} \ .
\label{eq:def_z}
\end{equation}
Defining the dimensionless wave-number as $q\!\equiv\!\tfrac{c_s D k}{|g'(1)| v_0}$, and recalling that we already defined above
the dimensionless (complex) growth rate as $\lambda\!\equiv\!\tfrac{\Lambda D}{|g'(1)|v_0}$,
Eq.~\eqref{eq:def_z} can be cast as $\lambda\!=\!-iqz$. With these definitions, Eq.~\eqref{eq:Hspectrum_dimensionless} can be rewritten as
\begin{equation}
s(z, q) \equiv \gamma \left(1 -i q z \right) \left[\sqrt{1-\beta^2 z^2}\, g_1(z, \beta) - i f g_2(z, \beta) \right]
- i z \left(\Delta -i q z \right) = 0 \ ,
\label{eq:spectrum_dimless}
\end{equation}
where
\begin{eqnarray}
g_1(z, \beta) = \frac{z^2}{1-\sqrt{1-z^2} \sqrt{1-\beta^2 z^2}} \qquad\hbox{and}\qquad g_2(z, \beta) = 2-g_1(z, \beta) \ .
\label{eq:g12_dimless}
\end{eqnarray}

As before, Eq.~\eqref{eq:spectrum_dimless} is an implicit expression for the explicit spectrum $z(q)$, which depends on the four dimensionless parameters
$\Delta$, $\beta$, $\gamma$ and $f$. Note also that due to algebraic manipulations,
Eq.~\eqref{eq:spectrum_dimless} no more follows the structure of Eq.~\eqref{eq:schematic_LSA}
(which is preserved in Eq.~\eqref{eq:spectrum_dim} and~\eqref{eq:Hspectrum_dimensionless}), rather the terms are mixed to some extent.
Later on, when discussing some of the physics behind our results, we will reinterpret them in terms of Eq.~\eqref{eq:schematic_LSA}.
Due to the appearance of the complex square root function in the above expressions, Eq.~\eqref{eq:spectrum_dimless} is understood
as having a branch-cut on the real axis along $|z|\!>\!1$ (there is also a branch-cut on the real axis along $|z|\!>\!1/\beta$ associated with $\sqrt{1-\beta^2 z^2}$.
Combinations of $\sqrt{1-z^2}$ and $\sqrt{1-\beta^2 z^2}$, as in Eq.~\eqref{eq:spectrum_dimless}, may have more complicated branch-cut structures).
The existence of these branch-cuts has implications that will be discussed later. Finally, in analyzing Eq.~\eqref{eq:spectrum_dimless} we assume that the dimensionless
wave-number $q$ --- and hence the dimensional wave-number $k$ ---  spans the whole interval $0\!<\!q\!<\!\infty$. The small wave-numbers limit, $q\!\to\!0$, is understood to imply $D k\!\ll\!|g'(1)|v_0/c_s$ while maintaining $k H\!\gg\!1$. This can always be guaranteed by having a sufficiently large $H$.

In the next parts of this section we present an extensive analysis of the linear stability spectrum in Eq.~\eqref{eq:spectrum_dimless}. As in Sect.~\ref{sec:finiteH}, we will establish relations between the unstable solutions of Eq.~\eqref{eq:spectrum_dimless} and various elastodynamic solutions and use this insight to derive analytic results that will shed light on the underlying physics. In Sect.~\ref{subsec:dilatational} we show that there exist unstable solutions related to dilatational waves propagating in the direction opposite to the sliding motion. In Sect.~\ref{subsec:shear} we show that there exist another class of unstable solutions which are related to shear waves propagating in the direction sliding motion. In Sect.~\ref{subsec:QS} we briefly review a qualitatively different class of solutions, which are not elastodynamic in nature, but rather quasi-static~\citep{Rice2001}. In Sect.~\ref{subsec:phase diagram} we present a comprehensive stability phase diagram, which puts together all three classes of solutions of the linear stability spectrum in Eq.~\eqref{eq:spectrum_dimless}. While we do not provide a mathematical proof that other classes of solutions do not exist, we suspect that our analysis is exhaustive.

\subsection{Dilatational wave dominated instability}
\label{subsec:dilatational}

In the spirit of Sect.~\ref{sec:finiteH}, we will look for solutions of Eq.~\eqref{eq:spectrum_dimless} that are related to propagating wave solutions. In particular, we note that the linear stability spectrum of Eq.~\eqref{eq:spectrum_dimless} significantly simplifies when $z\!=\!1/\beta$, for which $\sqrt{1-\beta^2 z^2}$ vanishes. Physically, the latter corresponds to $\delta\sigma_{xy}\!=\!0$, where $k_d(\Lambda,k)\!=\!0$ and $g_1(\Lambda,k)$ is finite
(cf. Eq.~\eqref{eq:pert_elastodynamics} with $g_1$ replacing $G_1$), i.e.~to frictionless boundary conditions. Substituting $z\!=\!1/\beta$ in Eq.~\eqref{eq:spectrum_dimless} and taking the limit $q\!\to\!0$, we immediately observe that it is a solution if $\gamma f (1/\beta^2-2)\!=\!\Delta/\beta$. As will be shown soon, the latter is an exact stability condition for the emergence of unstable solutions located near $z\!=\!1/\beta$ in the complex $z$-plane. Recall that a real $z$ is equivalent to $\Re(\Lambda)\!=\!0$, which is precisely where solutions change from growing ($\Re(\Lambda)\!>\!0$) to decaying ($\Re(\Lambda)\!<\!0$) in time.

$z\!=\!1/\beta$ corresponds to the dispersion relation for dilatational waves, $\Lambda\!=\!-i c_d k$ (i.e.~a frictionless boundary conditions,
$\delta\sigma_{xy}\!=\!0$), which means that instability modes located near $z\!=\!1/\beta$ in the complex-plane travel at nearly the dilatational wave-speed in the direction opposite to the sliding direction. The direction of propagation is a result of the minus sign in the last expression. It is important to stress in this context that while $z\!\to\!-1/\beta$ also corresponds to a dilatational wave ($\delta\sigma_{xy}\!=\!0$), it is {\em not} a solution of Eq.~\eqref{eq:spectrum_dimless}.
That is, friction in the presence of homogeneous sliding breaks the directional symmetry of dilatational waves.
The fact that $\Lambda$ vanishes in the limit $k\!\to\!0$, marks a crucial difference between the analysis to be performed here and
the one in Sect.~\ref{sec:finiteH}, where the finite system size $H$ implied a finite cutoff frequency $\Lambda\!\to\!\Lambda_0$ in the limit $k\!\to\!0$.

Following this simplified analysis, which indicates that some unstable solutions might be located near $z\!=\!1/\beta$ in
the complex $z$-plane, we aim at obtaining analytic results for the spectrum by a systematic expansion around this point.
That is, we are interested in obtaining a systematic expansion of the form $z\!=\!1/\beta + \delta z$, where $\delta z$ is a small complex number (i.e.~$|\delta z|\!\ll\!1$)
whose imaginary part determines the stability of sliding ($\Im(\delta z)\!>\!0$ implies instability and $\Im(\delta z)\!<\!0$ implies stability).
This should be done carefully, though, since $z\!=\!1/\beta$ is a branch-point, where a Laurent expansion does not exist.

To address this issue, let us briefly discuss one of the physical implications of being close to $z\!=\!1/\beta$.
First, note that the real part of $k_d$ in the elastodynamic solution in Eqs. \eqref{eq:elastic fields}
controls the decay length in the $y$-direction ($k_s$ plays a similar role, but is not discussed here. Note also that
$A_{3,4}\!\to\!0$ in the $H\!\to\!\infty$ limit considered here, which ensures the proper decay of solutions sufficiently away from the interface.).
Then, expressing $k_d$ in terms of $z$, $k_d\!=\!k\sqrt{1-\beta^2z^2}$, we observe that $k_d$ vanishes as $z\!\to\!1/\beta$, i.e. there is no decay in the $y$-direction in this case, and the proximity to $z\!=\!1/\beta$ actually controls the smallness of $k_d$ (for a given wave-number $k$). Therefore, we define a complex number $\kappa_d\!\equiv\!\sqrt{1-\beta^2 z^2}$ such that $k_d\!=\!k\,\kappa_d$, where $|\kappa_d|\!\ll\!1$.

With this definition of smallness, we go back to our original motivation to derive a systematic expansion around
$z\!=\!1/\beta$ and express $z$ in terms of $\kappa_d$ as
\begin{equation}
z=\frac{\sqrt{1-\kappa_d^2}}{\beta} \simeq 1/\beta - \frac{\kappa_d^2}{2\beta} + \mathcal{O}(\kappa_d^4) \ .
\label{eq:kappa_d}
\end{equation}
The latter expression has the desired form $z\!=\!1/\beta + \delta z$ and our next goal is to estimate $\kappa_d$ itself from the linear stability spectrum in Eq. \eqref{eq:spectrum_dimless}.
To do this, we need to rewrite the spectrum in terms of the new independent variable $\kappa_d$.
In the proximity of $z\!=\!1/\beta$, the functions $g_1(z)$ and $g_2(z)$ in Eq.~\eqref{eq:g12_dimless}
can be written in terms of $\kappa_d$ as follows
\begin{eqnarray}
\hat{g}_{1_\mp}(\kappa_d) \simeq \frac{1/\beta^2}{1\mp i \kappa_d \sqrt{1/\beta^2-1}} \quad\qquad\hbox{and}\quad\qquad \hat{g}_{2_\mp}(\kappa_d) \simeq 2-\hat{g}_{1_\mp}(\kappa_d) \ ,
\label{eq:g12_dimless1}
\end{eqnarray}
where the minus sign corresponds to the stable branch ($\Im(z)\!<\!0$) and the plus sign to the unstable branch ($\Im(z)\!>\!0$). This emerges from the limit $-\sqrt{1-z^2}\to \mp i \sqrt{1/\beta^2-1}$ as $z\!\to\!1/\beta$, where the different signs correspond to taking the limit from the two sides of the branch-cut ($\Im(\delta z)\!\to\!0^\mp$). The main advantage of Eqs. \eqref{eq:g12_dimless1} is that $\hat{g}_{1_\mp}(\kappa_d)$ and $\hat{g}_{2_\mp}(\kappa_d)$ are analytic such that a Laurent expansion around $\kappa_d\!=\!0$ exists\footnote{We note in passing that we could do the whole analysis with $\delta z$ instead of $\kappa_d$, invoking the leading term $\sim\!\sqrt{\delta z}$ in a fractional power series. The two routes are equivalent when identifying $\kappa_d\!\simeq\!\sqrt{-2\beta\,\delta z}$, which is precisely what Eq.~\eqref{eq:kappa_d} states.}.

Using Eqs.~\eqref{eq:g12_dimless1}, we can rewrite the linear stability spectrum in Eq. \eqref{eq:spectrum_dimless} in terms of $\kappa_d$ as
\begin{eqnarray}
s(\kappa_d, q) \simeq \gamma \left(1-i q/\beta \right) \left[\kappa_d\,\hat{g}_{1_\mp}(\kappa_d) - i f\,\hat{g}_{2_\mp}(\kappa_d) \right] - i/\beta \left(\Delta- i q/\beta \right) = 0 \ ,
\label{eq:spectrum_dimless1}
\end{eqnarray}
where we set $z\!=\!1/\beta$. We then linearize the following $\kappa_d$-dependent quantities
\begin{eqnarray}
k_d\,\hat{g}_{1_\mp}(\kappa_d) \!\simeq\! \kappa_d/\beta^2 + \mathcal{O}(\kappa_d^2),\quad\quad\quad \hat{g}_{2_\mp}(\kappa_d) \!\simeq\! 2 - 1/\beta^2\left(1 \pm i \kappa_d \sqrt{1/\beta^2-1}\right) + \mathcal{O}(\kappa_d^2) \ ,
\label{eq:kappad_expansion}
\end{eqnarray}
substitute them into Eq. \eqref{eq:spectrum_dimless1} and solve the resulting linear equation for $\kappa_d$, obtaining
\begin{equation}
\kappa_d \simeq \frac{i f \gamma \left(1-i q/\beta \right)\left(2-1/\beta^2 \right)+ i\left(\Delta - i q/\beta \right)\!/\beta}{\gamma\left(1- i q/\beta \right)\left(1 \mp f \sqrt{1/\beta^2 -1} \right)\!/\beta^2} \ .
\label{eq:kappa_d1}
\end{equation}
Substituting the latter in Eq. \eqref{eq:kappa_d}, we can calculate the dimensionless growth rate $\Re(\lambda)\!=\!q\Im(z)$ in the form
\begin{eqnarray}
\Re(\lambda) \simeq q^2 \left(1-\Delta \right) \left(\frac{[\gamma f \beta \left(1/\beta^2-2\right)-1] q^2/\beta^2 +\gamma f\beta(1/\beta^2-2)-\Delta}{\gamma^2\left(q^2/\beta^2+1\right)^2(1\mp f\sqrt{1/\beta^2-1})^2}\right) \ ,
\label{eq:analytic_spectrum}
\end{eqnarray}
where, as before, the stable solution corresponds to the minus sign and unstable one to the plus sign. This analytic prediction is one of the major results of this paper. It is important to stress that unlike the growth rate in Eq.~\eqref{eq:Hspectrum}, which was obtained by a small wave-numbers expansion, the growth rate in Eq.~\eqref{eq:analytic_spectrum} was obtained by an expansion in the complex plane near $z\!=\!1/\beta$. Consequently, it is valid --- as will be explicitly demonstrated below --- for {\em any} wave-number $q$.

A lot of analytic insight can be gained from Eq.~\eqref{eq:analytic_spectrum}. First, note that the growth rate $\Re(\lambda)$ in Eq.~\eqref{eq:analytic_spectrum} is continuous, but not differentiable at the transition
from the stable to unstable branches as a function of $q$ (i.e.~it has a kink due to the existence of a branch-cut in the equation for the spectrum). Then, we see that unstable modes appear in the long wavelength regime, $0\!<\!q\!<\!q_{_c}^{_{(d)}}$, where the critical wave-number $q_{_c}^{_{(d)}}$ is simply obtained from the condition $\Re(\lambda)\!=\!0$
\begin{equation}
q_{_c}^{_{(d)}} \simeq \beta \sqrt{\frac{\gamma f \beta \left(1/\beta^2-2 \right)-\Delta}{1-\gamma f \beta \left(1/\beta^2 -2 \right)}} \ .
\label{eq:qc}
\end{equation}
The instability threshold is obtained by taking the limit $q_{_c}^{_{(d)}}\!\to\!0$
(i.e.~the instability does not occur at a finite wavelength)
\begin{equation}
\gamma f = \frac{\Delta}{\beta\left(1/\beta^2-2\right)} \ ,
\label{eq:threshold}
\end{equation}
which is identical to the one derived at the beginning of this section.
When the left-hand-side is smaller than the right-hand-side, there exist no unstable modes,
i.e.~the regime $0\!<\!q\!<\!q_{_c}^{_{(d)}}$ shrinks to zero and $\Re(\lambda)$ is always negative.
In the opposite case, when the left-hand-side is larger than the right-hand-side, a finite range of unstable modes emerges.
A simple calculation shows that the threshold condition actually emerges from the numerator of $\Im[\kappa_d]$ in Eq.~\eqref{eq:kappa_d1}, and in particular from its $q$-independent part (since the threshold condition corresponds to the limit of vanishing wave-number $q$).

Let us discuss the physics embodied in Eq.~\eqref{eq:threshold}.
For that aim, recall the definitions of $\gamma$ and $\Delta$ in Eq.~\eqref{eq:def_dimless}
and substitute them in Eq.~\eqref{eq:threshold} to obtain
\begin{equation}
f \frac{\mu}{c_s} \beta\left(1/\beta^2-2\right)= \sigma_0\,d_v\!f \ .
\label{eq:threshold1}
\end{equation}
A first observation is that this instability threshold is independent of $\pa_v\!f$. Put differently,
as far as the threshold is concerned, the distinction between $d_v\!f$ and $\pa_v\!f$ is irrelevant
as if the friction law is only rate-dependent, $f(v)$. This can be understood as follows;
the right-hand-side of Eq. \eqref{eq:threshold1} corresponds to the $\sigma_0\,\delta f$ term (variation of the friction law)
in Eq.~\eqref{eq:schematic_LSA}. Near threshold we have $\Lambda\!\sim\!-i k\!\to\!0$,
which can be substituted in the expression for $\delta f$ in Eq. \eqref{eq:delta_f2}.
We observe that the term proportional to $\pa_v\!f$ scales as $k^2$, while the one proportional to
$d_v\!f$ scales as $k$ and hence the latter dominates the former.
Consequently, we have $\delta f\!\sim\! d_v\!f$, independently of $\pa_v\!f$.

Another aspect of Eq.~\eqref{eq:threshold1} which is worth noting concerns
the left-hand-side, which is proportional to $f$ and hence corresponds to the $f\delta\sigma_{yy}$ term
in Eq.~\eqref{eq:schematic_LSA}. Indeed, $\delta\sigma_{yy}$ in Eq. \eqref{eq:pert_elastodynamics} scales as $k$,
while $\delta\sigma_{xy}$ is higher order in $k$ and hence negligible.
Moreover, we observe that $\delta\sigma_{yy}$ is proportional to the so-called radiation damping factor for sliding~\citep{Rice1993, Rice2001, Crupi2013}, $\mu/c_s$, which essentially follows from dimensional considerations in the elastodynamic regime. To conclude, the present discussion shows  that Eq. \eqref{eq:threshold} is actually of the form $\sigma_0\,d_v\!f \!\sim\! f\,\delta\sigma_{yy}$, i.e.~the onset of instability is controlled by a balance between the stabilizing steady-state velocity-strengthening friction, $d_v\!f$, and the destabilizing elastodynamic bi-material effect, $\delta\sigma_{yy}$.

Next, we consider the analytic prediction for $\Re(\Lambda)$ in the limit $k\!\to\!\infty$
(or in dimensionless units, $\Re(\lambda)$ in the limit $q\!\to\!\infty$).
By counting powers in Eq. \eqref{eq:analytic_spectrum} we observe that in the limit $q\!\to\!\infty$,
$\Re(\lambda)$ approaches a constant whose sign is determined by the sign of $\gamma f \beta \left(1/\beta^2-2\right)-1$.
If $\gamma f \beta \left(1/\beta^2-2\right)\!<\!1$, then the constant is negative. This does not immediately imply
stability because, following the discussion above, a finite range of unstable modes emerges if
$\Delta\!<\!\gamma f \beta \left(1/\beta^2-2\right)\!<\!1$. If, on the other hand, we have
\begin{equation}
\gamma f \beta \left(1/\beta^2-2\right) > 1 \ ,
\label{eq:all_modes}
\end{equation}
then the constant is positive, which implies that {\em all} wave-numbers are unstable. Indeed, Eq.~\eqref{eq:qc} shows that the critical wave-number $q_{_c}^{_{(d)}}$ diverges in the limit $\gamma f \beta \left(1/\beta^2-2\right)\!\to\!1$.

We have thus seen that when elastodynamic effects become sufficiently strong, i.e.~when the combination $\gamma f$ becomes sufficiently large, all wave-numbers are unstable. This observation raises the issue of ill-posedness, which has been quite extensively discussed in the literature recently~\citep{Renardy1992, Adams1995, Martins1995a, Martins1995b, Simoes1998, Ranjith2001}. Ill-posedness is a stronger condition than instability for all wave-numbers,
i.e.~a problem can feature unstable modes at all wave-numbers but still be mathematically well-posed, and is defined as follows; consider the perturbation of any relevant interfacial field in the linear stability problem,
 e.g. the slip velocity field $\delta v$, and express it as an integral over all wave-numbers
\begin{equation}
\delta v(x,t) \sim \int_{-\infty}^\infty a(k) \exp[- i k x] \exp[\Lambda(k) t] dk \ ,
\label{eq:ill-posedness}
\end{equation}
where $a(k)$ is the amplitude of the $k$th mode.
If this integral fails to converge, the problem is regarded as mathematically ill-posed.

An important example in this context~\citep{Renardy1992, Adams1995, Martins1995a, Martins1995b, Simoes1998, Ranjith2001} is sliding along a bi-material interface described by Coulomb friction,
$\tau\!=\!\sigma f$ (where $f$ is a constant). In this case, $\Re(\Lambda(k))\!\sim\!|k|$ (with a positive prefactor)
and the integral in Eq.~\eqref{eq:ill-posedness} fails to converge for any $x\!\ne\!0$ at any finite time,
unless $a(k)$ decays exponentially or stronger with $|k|$. The problem can be made well-posed if in response to normal stress variations, $\tau\!=\!\sigma f$ is approached over a finite time scale~\citep{Ranjith2001}.
In our problem, within the standard rate-and-state friction framework, we saw above that there exists a range of parameters in which all wave-numbers are unstable. Yet, in this case $\Re(\Lambda(k))$ approaches a constant as $k\!\to\!\infty$, in which case the integral in Eq.~\eqref{eq:ill-posedness} converges. Therefore, we conclude that the response of bi-material interfaces described by standard rate-and-state friction laws is mathematically well-posed.

\begin{figure}[here]
 \centering
 \begin{tabular}{ccc}
\includegraphics[width=0.5\textwidth]{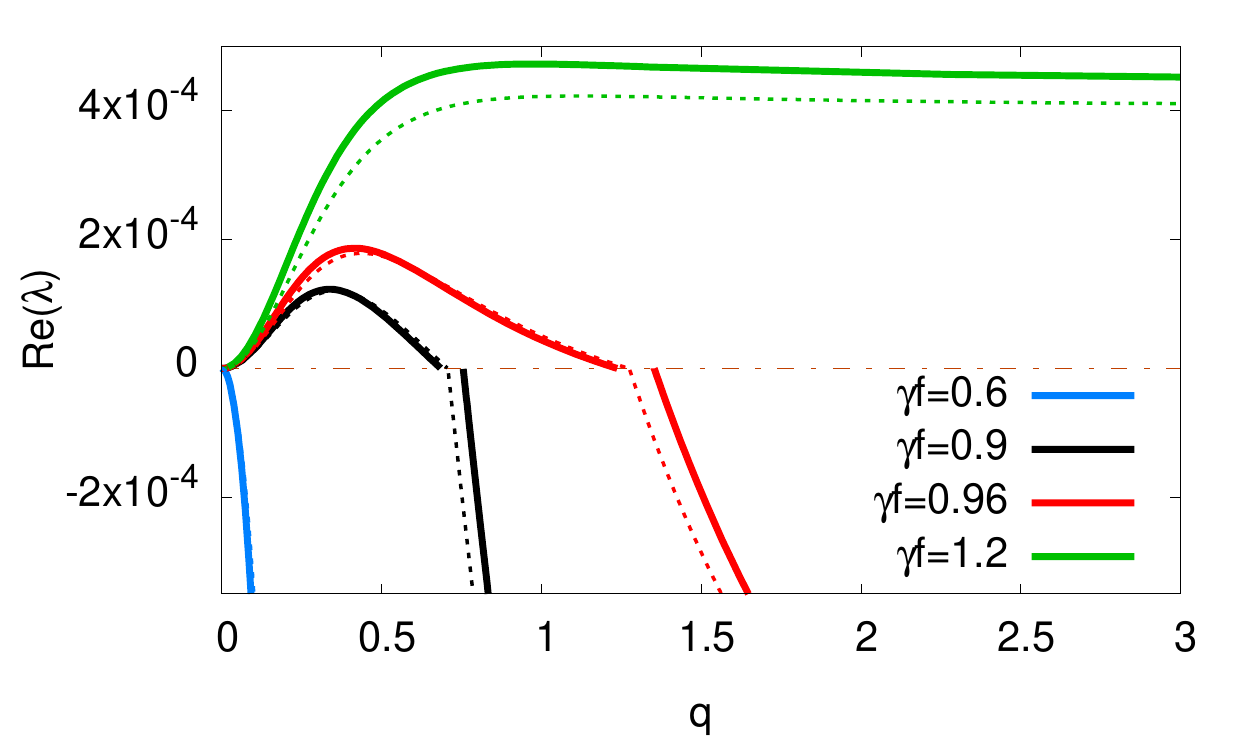}
\includegraphics[width=0.5\textwidth]{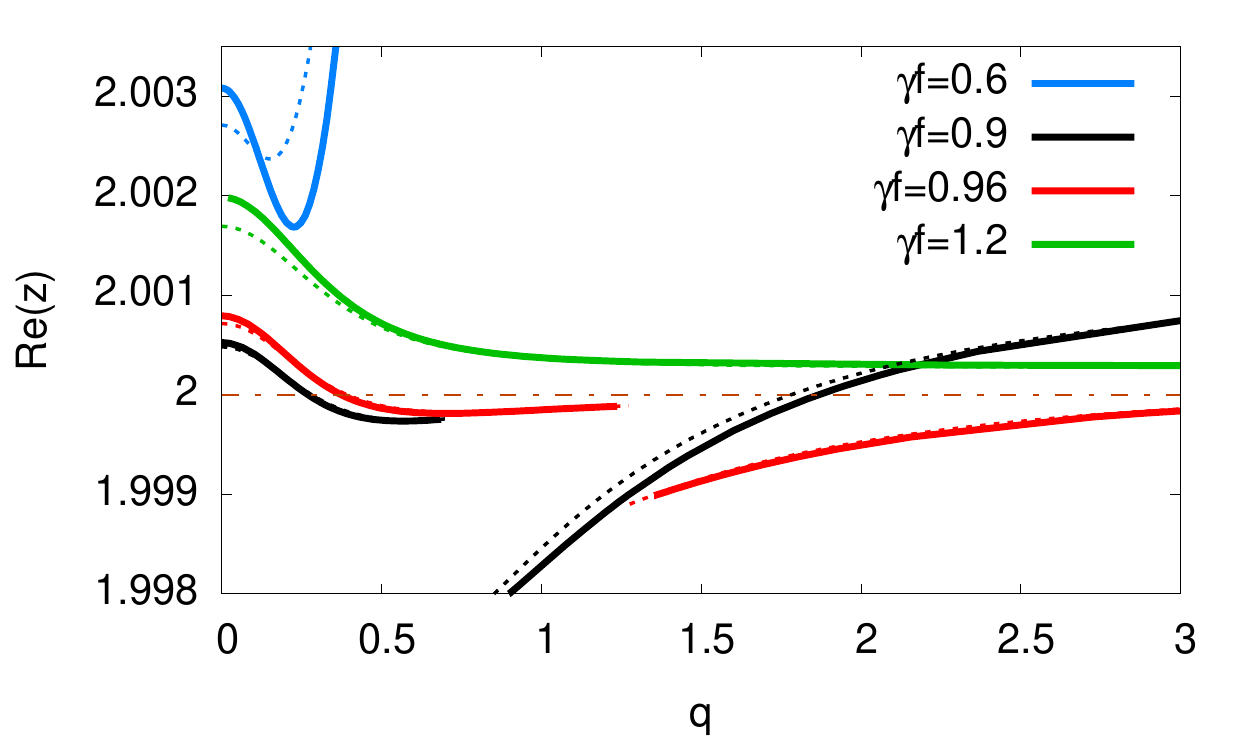}
 \end{tabular}
 \caption{(left) The dimensionless growth rate $\Re(\lambda)\!=\!q \Im(z)$ vs. the dimensionless wave-number $q$ for various $\gamma f$'s ($\gamma f$ was changed by changing $\gamma$). The solid lines show the numerical solution of the linear stability spectrum in Eq.~\eqref{eq:spectrum_dimless} for both the stable ($\Re(\lambda)\!<\!0$) and unstable ($\Re(\lambda)\!>\!0$, when it exists) branches. The dotted lines correspond to the analytical prediction of Eq.~\eqref{eq:analytic_spectrum}.
 A very good quantitative agreement between the analytic prediction and the direct numerical solution is demonstrated (see text for more details). The discontinuities (gaps) observed in the full numerical solution are discussed in Appendix~\ref{appendix_gap}. (right) $\Re(z)$ vs. the wave-number $q$. The parameters used are $f\!=\!0.3,\ \beta\!=\!0.5$ and $\Delta\!=\!0.7$.}
 \label{fig:comparison}
\end{figure}

We are now in a position to quantitatively compare the analytic predictions derived from Eq. \eqref{eq:analytic_spectrum}
to a direct numerical solution of the linear stability spectrum in Eq.~\eqref{eq:spectrum_dimless}. The results are shown in Fig.~\ref{fig:comparison}. On the left panel, $\Re(\lambda)\!=\!q\,\Im(z)$ is shown as a function of $q$ for various $\gamma f$'s and fixed representative values of $\Delta$, $f$ and $\beta$.

All in all, Fig.~\ref{fig:comparison} demonstrates a good quantitative agreement between the analytic prediction and the full numerical solution over a significant range of parameters and wave-numbers. In particular, Eq. \eqref{eq:threshold} predicts (for the chosen $\beta$) that the onset of instability takes place at $\gamma f\!=\!0.7$, which is precisely what is observed. Furthermore, the onset of instability appears at $k\!\to\!0$, as predicted. The critical wave-number $q_{_c}^{_{(d)}}$, predicted in Eq. \eqref{eq:qc}, is quantitatively verified for several sets of parameters above threshold. Finally, the shape of the unstable spectrum, including the constant asymptote as $q\!\to\!\infty$, is quantitatively verified.

The only interesting deviation of the analytic prediction of Eq.~\eqref{eq:analytic_spectrum}
from the full numerical solution in the left panel of Fig.~\ref{fig:comparison} is that the latter exhibits
a discontinuity (a gap) at the transition from the unstable to the stable part of the solution,
while the former is continuous but rather exhibits a discontinuous derivative.
This results in a shift of the stable part of the spectrum when an unstable range of wave-numbers exists.
The origin of the gap in the spectrum is explained in Appendix~\ref{appendix_gap}.
On the right panel of Fig.~\ref{fig:comparison}, $\Re(z)$ is shown as a
function of $q$ for both the numerical solution of Eq.~\eqref{eq:spectrum_dimless}
and the real part of Eq.~\eqref{eq:kappa_d} (together with Eq.~\eqref{eq:kappa_d1}).
The figure demonstrates, again, a good quantitative agreement between the analytic prediction and the exact numerical solution. Furthermore, the two panels of Fig.~\ref{fig:comparison} show that indeed $\Im(z)\!=\!\Re(\lambda)/q\!\ll\!\Re(z)\!\simeq\!1/\beta$, as expected for solutions located near $z\!=\!1/\beta$. In particular, note that solutions remain close to $z\!=\!1/\beta$ in the complex-plane for {\em every} wave-number in this class of solutions.

With this we complete the discussion of the dilatational wave dominated instability, which corresponds to unstable modes of predominantly dilatational wave nature propagating in the direction opposite to the sliding direction (corresponding to solutions near $z\!=\!1/\beta$).
In the next subsection we discuss a distinct class of unstable solutions of the linear stability spectrum in Eq.~\eqref{eq:spectrum_dimless}.

\subsection{Shear wave dominated instability}
\label{subsec:shear}

Inspired by the discussion in the previous subsection, we look here for another class of unstable solutions. This time we focus on the zeros of $\sqrt{1-z^2}$, in particular on solutions located near $z\!=\!-1$ in the complex-plane. As will be shown below, these instability modes are of shear wave-like nature, propagating with a phase velocity close to $c_s$ in the sliding direction.

\begin{figure}
\begin{center}
\includegraphics[width=12cm]{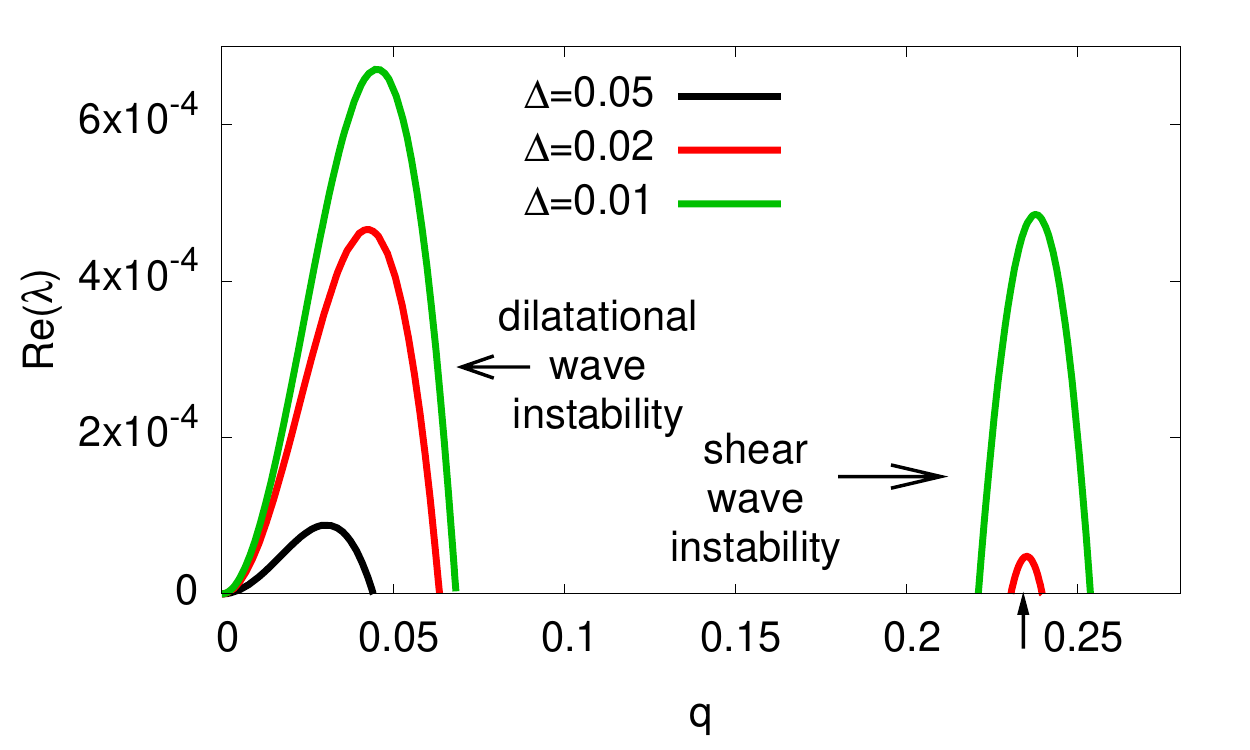}
\caption{$\Re(\lambda)$ vs. wave-number $q$ for both the dilatational and shear wave dominated instabilities for various $\Delta$'s, and $\gamma\!=\!0.25$. For the parameters used (together with $f\!=\!0.3$ and $\beta\!=\!0.5$), Eqs.~\eqref{eq:full_shear_inst}-\eqref{eq:Delta_m} imply $q_{_c}^{_{(s)}}\!\simeq\!0.234$ (marked with a vertical arrow) and $\Delta^{(m)}\!\simeq\!0.024$. We observe that for $\Delta\!<\!\Delta^{(m)}$ the two types of instability coexist, while for $\Delta\!>\!\Delta^{(m)}$ only the dilatational one exists. Furthermore, the figure quantitatively verifies the prediction for the critical wave-number $q_{_c}^{_{(s)}}$, demonstrating that indeed the shear wave dominated instability appears at a finite wave-number, unlike the dilatational wave one. The growth rate of the dilatational wave-like instability is larger than the one corresponding to the shear wave-like instability.}
\label{fig:both}
\end{center}
\end{figure}

To see how this rigorously emerges, we set $z\!=\!-1$ in Eq. \eqref{eq:spectrum_dimless} (which corresponds to $\Re(\Lambda)\!=\!0$, i.e.~to the threshold of instability) and separate the real and imaginary parts to obtain
\begin{equation}
q_{_c}^{_{(s)}} = \frac{\gamma\sqrt{1-\beta^2}}{1-\gamma f}  \quad\qquad\hbox{and}\quad\qquad  \Delta = \gamma f - \frac{(\gamma f)^2(1-\beta^2)}{(1-\gamma f)f^2} \ ,
\label{eq:full_shear_inst}
\end{equation}
where $q_{_c}^{_{(s)}}$ is the critical (dimensionless) wave-number at threshold and
the second relation is the onset of instability condition (an instability occurs when the left-hand-side is smaller than the right-hand-side). This is an exact result. Unlike the dilatational wave dominated instability, which featured a vanishing critical wave-number at threshold, $q_{_c}^{_{(d)}}\!\to\!0$, the shear wave dominated instability takes place at a finite wave-number (above threshold, a finite range of unstable $q$'s emerges around $q_{_c}^{_{(s)}}$, cf. Fig.~\ref{fig:both}). $z\!=\!-1$ corresponds to the dispersion relation for shear waves, $\Lambda\!=\!i c_s k$, which means that this instability is mediated by modes propagating at nearly the shear wave-speed in the direction of sliding. The propagation direction is determined by the positive sign in the last expression. It is important to stress in this context that $z\!=\!1$ is not a solution of Eq.~\eqref{eq:spectrum_dimless}, again demonstrating the symmetry breaking induced by frictional sliding.

The dilatational wave dominated instability exists for all physically relevant values of $\Delta$, i.e.~for $0\!\le\!\Delta\!\le\!1$. Is it true also for the shear wave dominated instability? To address this question, we interpret $\Delta$ in Eq.~\eqref{eq:full_shear_inst} as a function of $\Gamma\!\equiv\!\gamma f$, parameterized by $\beta$ and $f$. $\Delta(\Gamma)$ is a non-monotonic function which attains a maximum at
\begin{equation}
\Delta^{(m)} = \frac{2(1-\beta^2) + f^2 -2 \sqrt{(1-\beta^2)(1-\beta^2 + f^2)}}{f^2} \ .
\label{eq:Delta_m}
\end{equation}
For realistic values of the friction coefficient $f$ (i.e.~$f\!\sim\!0.2\!-\!0.75$), we have $\Delta^{(m)}\!\ll\!1$,
which shows that the shear wave dominated instability is characterized by a small $\Delta$.
Furthermore, $\Delta(\Gamma)$ in Eq.~\eqref{eq:full_shear_inst} vanishes at $\Gamma\!=\!f^2/(1-\beta^2+f^2)$, which is also typically small due to the smallness of $f^2$. In fact, if we assume a small $\Gamma$ and invoke a parabolic approximation
for $\Delta(\Gamma)$ in Eq. \eqref{eq:full_shear_inst}, we obtain for the maximum
$\Delta^{(m)}\!\simeq\!f^2/[4(1-\beta^2)]$, which is just the leading contribution in the expansion
of Eq.~\eqref{eq:Delta_m} in terms of $f^2$. We thus conclude that the shear wave dominated instability is localized
in a relatively small region near the origin in the $\Delta\!-\!\Gamma$ plane.

One implication of the above discussion is that since in the stability boundary of the dilatational wave dominated instability $\Delta$ increases linearly with $\Gamma\!=\!\gamma f$, cf. Eq.~\eqref{eq:threshold}, the shear and dilatational waves instabilities coexist only in a relatively small range of $\Delta$'s, $0\!<\!\Delta\!<\!\Delta^{(m)}$.
To explicitly demonstrate this, $\Re(\lambda)$ is plotted vs. $q$ in Fig.~\ref{fig:both} for the two types of instabilities
and various small $\Delta$'s. We observe that indeed the two instabilities coexist for $0\!<\!\Delta\!<\!\Delta^{(m)}$,
but only the dilatational one exists for $\Delta\!>\!\Delta^{(m)}$ (see figure caption for details), and that the growth rate of the dilatational instability is larger than that of the shear one. Furthermore, we see that indeed the dilatational wave dominated instability appears at a vanishing wave-number, while the shear wave dominated instability appears at a finite wave-number. The results for the shear wave dominated instability presented in Fig.~\ref{fig:both} were obtained numerically. We could have followed a similar procedure to the one taken in great detail in Sect.~\ref{subsec:dilatational} and derive analytic results by systematically expanding around $z\!=\!-1$ in the complex-plane. In order not to further complicate the presentation, we do not present this analysis here, but rather present numerical demonstrations of the main physical points.

The analysis of the spectrum in the large $k H$ limit presented so far has revealed two classes of elastodynamic-controlled unstable modes, one mediated by dilatational wave-like modes propagating in the direction opposite to the sliding motion
(corresponding to solutions near $z\!=\!1/\beta$) and one mediated by shear wave-like modes propagating in the direction of sliding (corresponding to solutions near $z\!=\!-1$). Related observations on the directionality of unstable modes and rupture along bi-material frictional interfaces have been previously made, see for example~\citet{Ranjith2001, Cochard2000, Adams2000, Weertman1980, Andrews1997, Ben-Zion2002, Adams1995, Adams1998,  Harris1997, Xia2004, Ampuero2008a}.

Finally, we emphasize that one cannot naturally superimpose the results of Eq.~\eqref{eq:Hspectrum} and Fig.~\ref{fig:finiteH} on those appearing in Fig.~\ref{fig:both} in a generic manner because the former results depend on $H$, while the latter do not (i.e. they are valid for $k\!\gg\!H^{-1}$). In particular, the growth rate in Fig.~\ref{fig:finiteH} decays as $H^{-1}$ and its relative magnitude compared to the growth rates in Fig.~\ref{fig:both} depends on the value of $H$. It is important to note, though, that for a real system with a given $H$ our results allow one to calculate the growth rate of all of the instabilities discussed above and determine which is the largest.

\subsection{The quasi-static limit}
\label{subsec:QS}

Up to now we have found two classes of elastodynamic-controlled instabilities, one related to dilatational waves and one to shear waves. In addition to these, there exists also a quasi-static class of unstable modes at extremely small $\Delta$'s, which is qualitatively different as it is not elastodynamic in nature.
This quasi-static instability has been discussed quite extensively in~\citet{Rice2001}, where the analysis has been performed for general bi-material interfaces (i.e.~not only for a strong contrast). Our goal here is just to briefly summarize those  results of~\citet{Rice2001} which are relevant to our discussion. In order to see how the quasi-static limit emerges, we need to take the limit of large wave-speeds $c_{s,d}\!\to\!\infty$ in the linear stability spectrum in Eq.~\eqref{eq:spectrum_dim}, while keeping their ratio $\beta$ fixed\footnote{Note that if the limit $c_{s,d}\!\to\!\infty$ is taken at the level of the fields themselves in Eq.~\eqref{eq:elastic fields}, the solutions become degenerate and additional independent solutions should be included.}. Obviously, the friction part remains unaffected and the elasticity parts are changed according to
\begin{equation}
k_d \to |k|, \quad\quad g_1 \to \frac{4(1-\nu)}{3-4\nu}, \quad\quad g_2 \to \frac{2(1-2\nu)}{3-4\nu} \ .
\end{equation}

The resulting quasi-static linear stability spectrum can be analyzed following~\citet{Rice2001}, leading to the stability condition
\begin{equation}
\Delta \simeq \tfrac{1}{4} f^2 \beta^4 + {\cal O}\left(\Gamma^2\right) \ ,
\label{eq:riceQS}
\end{equation}
where an instability appears when the left-hand-side is smaller than the right-hand-side\footnote{Equation~\eqref{eq:riceQS} coincides with the last equation on page 1890 of~\citet{Rice2001} (the equations in that paper are not numbered). Note, however, that $\beta$ in~\citet{Rice2001} denotes the Dundurs parameter (which is a function of the $4$ linear isotropic elastic moduli of the two sliding bodies) and is different from our $\beta$. In the limit of infinite material contrast, which is the limit we consider, the Dundurs parameter simply equals $\sqrt{\beta}$ (where $\beta$ is defined in Eq.~\eqref{eq:beta}). One should bear this in mind when comparing Eq.~\eqref{eq:riceQS} to the results of~\citet{Rice2001}.}. We do not report here on the detailed calculations that show that the leading elastodynamic effect on this instability branch enters only to quadratic order in $\Gamma\!=\!\gamma f$ (which, as discussed above, quantifies the importance of elastodynamic effects) and we do not explore the range of existence of this instability branch with increasing $\Gamma$.
For our purposes here it would be sufficient to note that to leading order, the instability condition in Eq.~\eqref{eq:riceQS} is independent of $\Gamma$. The quasi-static instability emerges at a finite wave-number (again, this is not shown explicitly here), $q_{_c}^{_{(qs)}}\!>\!0$ and consequently, a finite range of unstable modes exists above the threshold.

The quasi-static instability exists at extremely small values of $\Delta$, typically of the order of $10^{-3}$ due to the appearance of higher powers of $f$ and $\beta$ (both smaller than unity) in Eq.~\eqref{eq:riceQS}. Yet, since the stability condition for both the dilatational and shear wave dominated instabilities satisfies $\Delta\!\propto\!\Gamma$ for sufficiently small $\Gamma$, there exists a small region near the origin of the $\Delta\!-\!\Gamma$ plane, where only a quasi-static instability can be found.
All of these issues will be addressed next, where we construct the stability phase diagram of the problem
in the large $k H$ limit.

\subsection{Stability phase diagram}
\label{subsec:phase diagram}

One of the hallmarks of a linear stability analysis is a stability phase diagram in the space of the relevant physical parameters. The detailed analysis presented above allows us at this point to analytically construct such a stability phase diagram. We have identified three classes of instabilities in the large $k H$ limit (a dilatational wave dominated instability, a shear wave dominated instability and a quasi-static instability) and the corresponding stability boundaries are given in Eqs.~\eqref{eq:threshold}, \eqref{eq:full_shear_inst} and \eqref{eq:riceQS}. We rewrite these as
\begin{equation}
\Delta_d(\Gamma) = \Gamma \beta\left(1/\beta^2-2\right), \qquad \Delta_s(\Gamma) = \Gamma - \frac{\Gamma^2(1-\beta^2)}{(1-\Gamma)f^2}, \qquad \Delta_{qs}(\Gamma) \simeq \tfrac{1}{4} f^2 \beta^4 + {\cal O}\left(\Gamma^2\right) \ ,
\label{eq:stability_boundaries}
\end{equation}
where we added the subscripts $d$, $s$ and $qs$ to correspond to ``dilatational'', ``shear'' and ``quasi-static'' instabilities, respectively.

We treat all of these (in)stability boundaries as functions of $\Gamma\!=\!\gamma f$, parameterized by $f$ and $\beta$.
While there is some degree of arbitrariness in this choice, we strongly believe that it is the most natural
way to represent the interplay between the various physical effects in the problem,
within the framework of a two-dimensional phase diagram. An instability is implied whenever
$\Delta$ is below at least one of the $\Delta_d(\Gamma)$, $\Delta_s(\Gamma)$ or $\Delta_{qs}(\Gamma)$ lines
in the $\Delta\!-\!\Gamma$ plane. Note that while the results for $\Delta_d$ and $\Delta_s$ are exact, the one for $\Delta_{qs}$ is given to leading order in $\Gamma$. This will be enough for our purposes here. Finally, we stress that the phase diagram to be presented and discussed below pertains to the large $k H$ limit; in the limit of small $k H$, as discussed in Sect.~\ref{sec:finiteH}, homogeneous sliding is unconditionally unstable.

\begin{figure}[here]
 \centering
 \begin{tabular}{ccc}
\includegraphics[width=0.98\textwidth]{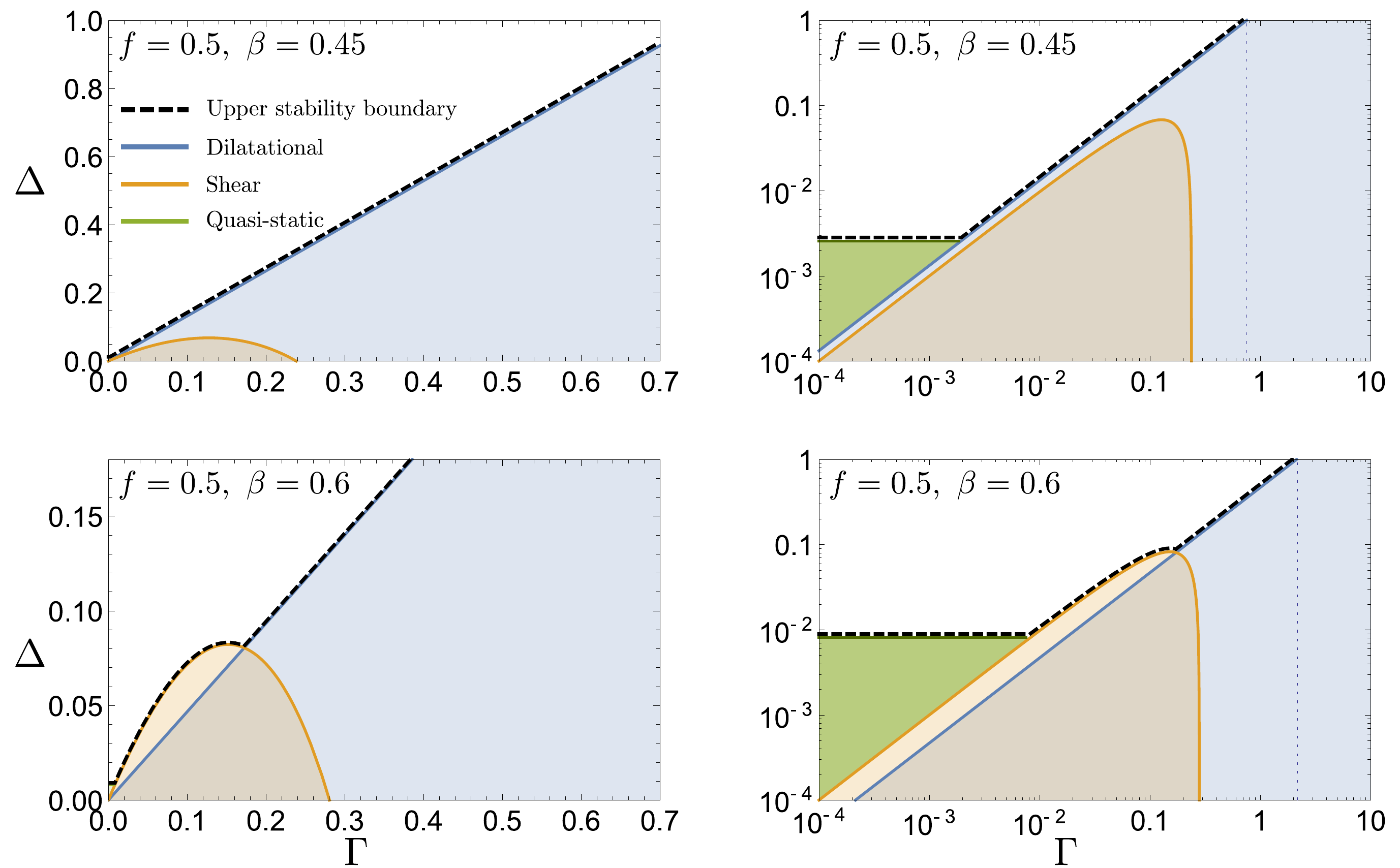}
 \end{tabular}
 \caption{Stability phase diagram in the $\Delta\!-\!\Gamma$ plane. Each row presents the same data (corresponding to the same $f$ and $\beta$), but the left column is in linear coordinates and the right one in logarithmic coordinates. Sliding is unstable beneath the upper stability boundary (dashed black line), which is the upper envelope of the three stability lines of Eq.~\eqref{eq:stability_boundaries}. The shaded colored areas correspond to regions of the phase diagram in which one or more instabilities exist (i.e. below the upper stability boundary, where each color corresponds to a different instability). In the two right panels, all dilatational modes are unstable in the region to the right of the dotted line, see Eq.~\eqref{eq:all_modes}. The quasi-static instability exists in a very small region near the origin of the $\Delta\!-\!\Gamma$ plane and hence is seen only in the logarithmic plots (right panels). Note also that since we only consider the quasi-static stability boundary to leading order in $\Gamma$, we truncated it at small values of $\Gamma$. For concreteness, we chose to truncate it when it intersects one of the other stability boundaries, though there is nothing physically special about these points (beyond the fact that they belong to the upper stability limit).}
 \label{fig:stability_diagram}
\end{figure}

As there exist various types of instabilities, one is interested in understanding under what conditions they
coexist and what is the upper stability boundary, i.e.~the line which separates the region in parameter space
where no instabilities exist at all from the region in which at least one instability exists.
In Fig.~\ref{fig:stability_diagram}, the $\Delta\!-\!\Gamma$ stability phase diagram for two sets of values of $(f, \beta)$
is shown. Note that negative values of $\Delta$, which correspond to steady-state velocity-weakening,
are not shown as in this case there is an instability independently of other parameters.

Let us discuss in detail the linear stability phase diagram. $\Delta_d(\Gamma)$ is a straight line spanning the whole range of $\Delta$ values, $0\!\le\!\Delta\!\le\!1$, while the other lines are localized near the origin
(readers are advised to remind themselves the physical meaning of the dimensionless parameters used here,
as discussed around Eq.~\eqref{eq:def_dimless}). In this sense, the dilatational wave dominated instability is
the main instability mode of the system in the large $k H$ limit. As discussed in Sect.~\ref{subsec:dilatational}, the trends in $\Delta_d(\Gamma)$ are clear; for a fixed $\Gamma$, increasing $\Delta$ (essentially increasing $d_v\!f$, recall that this stability boundary is independent of the direct effect $\pa_v\!f$, as discussed around Eq. \eqref{eq:threshold1}) promotes stability. Alternatively, for a fixed $\Delta$, instability is promoted by increasing $\Gamma$ (either by enhancing the elastodynamic bi-material effect $\sim\!\mu/c_s$ or by decreasing the normal load $\sigma_0$).

$\Delta_s(\Gamma)$, as discussed in Sect.~\ref{subsec:shear}, is a non-monotonic function that is bounded in the region
$0\!\le\!\Delta\!\le\!\Delta^{(m)}$ and crosses zero at a finite small $\Gamma$, $\Gamma\!=\!f^2/(1-\beta^2+f^2)$
(recall that $\Delta^{(m)}\!\ll\!1$ is defined in Eq.~\eqref{eq:Delta_m}).
This implies that, since $\Delta_d(\Gamma)$ is a straight line starting at the origin,
the shear and dilatational wave dominated instabilities always have a range of coexistence.
The question then is whether there exists a parameter range where the shear wave instability exists and
the dilatational one does not. It is evident from Eq.~\eqref{eq:stability_boundaries} that this depends on
the value of $\beta$. For small $\Gamma$, we have $\Delta_s(\Gamma)\!\simeq\! \Gamma + {\cal O}\left(\Gamma^2\right)$,
i.e.~$\Delta_s(\Gamma)$ starts linearly with a unit slope. $\Delta_d(\Gamma)$ is always linear with
a slope $\beta\left(1/\beta^2-2\right)$. Therefore, for $\beta\left(1/\beta^2-2\right)\!\ge\!1$ the shear wave dominated instability coexist with the dilatational one, while for $\beta\left(1/\beta^2-2\right)\!<\!1$, there exists a region above the $\Delta_d(\Gamma)$ line
in which the shear wave dominated instability exists, but the dilatational one does not. In the upper panels of Fig.~\ref{fig:stability_diagram} we used $\beta\!=\!0.45$, which is an example of the former, while in the lower panels we used $\beta\!=\!0.6$, which is an example of the latter.

$\Delta_{qs}(\Gamma)$ is, to leading order, a positive constant independent of $\Gamma$.
Since both $\Delta_d(\Gamma)$ and $\Delta_s(\Gamma)$ vanish linearly at small $\Gamma$,
there always exists an instability region controlled by quasi-static modes.
This quasi-static stability boundary joins either the $\Delta_d(\Gamma)$ line or the $\Delta_s(\Gamma)$ line,
depending on the value of $\beta$. It happens at $\Gamma\!\simeq\!f^2\beta^3/[4(1/\beta^2-2)]$ in the former case
and at $\Gamma\!\simeq\!f^2\beta^4/4$ in the latter. The two possibilities are shown on the two right panels in
Fig.~\ref{fig:stability_diagram}. The upper stability boundary always starts with $\Delta_{qs}(\Gamma)$,
and then either merges directly with $\Delta_d(\Gamma)$ (if $\beta\left(1/\beta^2-2\right)\!\ge\!1$)
or first merges with $\Delta_s(\Gamma)$ (if $\beta\left(1/\beta^2-2\right)\!<\!1$) which then merges with
$\Delta_d(\Gamma)$. One way or the other, except for a small region near the origin of the $\Delta\!-\!\Gamma$ plane,
the upper stability boundary is determined by $\Delta_d(\Gamma)$. As mentioned above, we do not consider higher order corrections to $\Delta_{qs}(\Gamma)$ in terms of $\Gamma$ and hence we truncate the latter when it intersects either $\Delta_s(\Gamma)$ or $\Delta_d(\Gamma)$ in Fig.~\ref{fig:stability_diagram} (we stress, though, that there is nothing special about the intersection point from the perspective of $\Delta_{qs}(\Gamma)$). Finally, note that in a region of coexistence, the instability with the largest growth rate $\Re(\Lambda)$ will be observed physically, cf. Fig.~\ref{fig:both}.

With this we complete the discussion of the linear stability analysis in the large $k H$ limit for large contrast
bi-material interfaces described by standard rate-and-state friction. In what follows,
we will discussed some generalized rate-and-state friction models,
which incorporate a modified non-instantaneous response to normal stress variations.

\section{Generalized friction models: Modified response to normal stress variations}
\label{sec:modified-models}

Up to now we considered the standard rate-and-state friction model in which the frictional resistance does not exhibit
a finite time response to variations in the local normal stress. That is, the friction stress $\tau(x,t)$ was affected
by the local interfacial normal stress $\sigma(x,t)$ only through Eq.~\eqref{eq:rsf_stress}, and the response was instantaneous. Nevertheless, some experimental work~\citep{Linker1992, Dieterich1992a, Prakash1992, Prakash1993, Prakash1998, Richardson1999, Bureau2000} indicated that the frictional resistance may exhibit a finite time response to normal stress variations. To take this possibility into account we adopt here two experimentally-based modifications of the constitutive relation and incorporate them into a generalized analysis. The goal is to understand the physical effects of the modified constitutive relations on the stability analysis.

The first modification we consider is due to~\citet{Prakash1992, Prakash1993, Prakash1998} and amounts to taking the time scale $T$ in Eq.~\eqref{eq:gradual_sigma} to be finite.
A finite $T$ means that when the interfacial normal stress $\sigma(x,t)$ varies, the friction stress $\tau(x,t)$ does not immediately follow it as in Eq.~\eqref{eq:rsf_stress}. $T$ is commonly expressed as $T\!=\!\alpha\!_{_{PC}}D/v$, where $\alpha\!_{_{PC}}$ is a dimensionless and a positive parameter which measures $T$ in units of the already existing time scale in the model, $D/v$. Consequently, $F(\tau,\sigma,v,\phi)$ in Eq. \eqref{eq:constitutive} takes the form
\begin{equation}
\dot\tau = F(\tau,\sigma,v,\phi) = -\frac{v}{\alpha\!_{_{PC}}D}\left(\tau - \sigma f(\phi,v)\right) \ .
\label{eq:gradual_sigma1}
\end{equation}

The instantaneous response in Eq. \eqref{eq:rsf_stress} is recovered in the limit $\alpha\!_{_{PC}}\!\to\!0$.
For any $\alpha\!_{_{PC}}\!>\!0$, the response is gradual such that the larger $\alpha\!_{_{PC}}$ the slower the response.
As the variation of the interfacial normal stress $\sigma(x,t)$ with slip (the elastodynamic bi-material effect)
is the major destabilizing effect in the analysis presented up to now and since $\alpha\!_{_{PC}}\!>\!0$
delays the destabilizing effect on the friction stress $\tau(x,t)$, we expect $\alpha\!_{_{PC}}\!>\!0$
to promote stability. The finite-time response to normal stress variations in Eq.~\eqref{eq:gradual_sigma1} has been invoked by several authors in relation to the regularization of the Coulomb friction law, for example in the context of the stability of steady frictional sliding~\citep{Ranjith2001}, which was already mentioned above, and dynamic rupture propagation~\citep{Cochard2000, Ben-Zion2002} along bi-material interfaces.

The second modification we consider is due to Linker and Dieterich~\citep{Linker1992, Dieterich1992a},
who interpreted their experiments as suggesting that the state variable $\phi$
is directly affected by variations in the normal stress.
In particular, they proposed that $G(\tau,\sigma,v,\phi)$ in Eq. \eqref{eq:constitutive} takes the form
\begin{equation}
\dot\phi = G(\tau,\sigma,v,\phi) = 1 - \frac{v\,\phi}{D} - \alpha\!_{_{LD}} \frac{\dot\sigma}{\sigma \pa_\phi f}\ ,
\label{eq:rsf_phi1}
\end{equation}
where $\alpha\!_{_{LD}}$ is a positive dimensionless parameter. When $\alpha\!_{_{LD}}\!=\!0$,
Eq.~\eqref{eq:rsf_phi} is recovered with $g(x)\!=\!1-x$ (note that $g'(1)\!=\!-1$).
To qualitatively understand the effect of $\alpha\!_{_{LD}}\!>\!0$ on the frictional stability, we consider a fast variation in the normal stress (i.e.~a large $|\dot\sigma|$) such that the last term on the right-hand-side of Eq.~\eqref{eq:rsf_phi1} dominates the other two terms. In this case, we can eliminate the time derivative to obtain
\begin{equation}
\delta\phi \simeq -  \frac{\alpha\!_{_{LD}}\delta\sigma}{\sigma_0 \pa_\phi f} \quad\quad \Longrightarrow \quad\quad
 \sigma_0\delta f \simeq -\alpha\!_{_{LD}}\delta\sigma \ ,
\label{eq:LD_simple}
\end{equation}
where we set $\sigma\!=\!\sigma_0$.

The last result may appear somewhat counterintuitive. To better understand it,
note that for $\alpha\!_{_{PC}}\!=\!0$ (which is assumed here to simplify things),
$\sigma_0\delta f$ contributes to the variation of the friction stress $\delta\tau$ as in
Eq.~\eqref{eq:schematic_LSA}. According to Eq. \eqref{eq:LD_simple}, when the normal stress reduces,
$\dot\sigma\!<\!0$, we have $\sigma_0\delta f\!>\!0$ which means the latter makes a {\em positive}
contribution to the friction stress (for $\alpha\!_{_{LD}}\!>\!0$).
This appears to suggest that the bi-material effect in this case is stabilizing. This is not quite the case,
because the effect of $\delta\sigma$ on $\delta\tau$ contains in fact also the last term on the right-hand-side
of Eq.~\eqref{eq:schematic_LSA}. Together, we obtain
$\delta\tau\!\simeq\!(f-\alpha\!_{_{LD}})\delta\sigma$ (recall that $\delta\sigma_{yy}\!=\!-\delta\sigma$), which
shows that as long as $f\!>\!\alpha\!_{_{LD}}$, a reduction in the normal stress still leads to a reduction in $\delta\tau$,
i.e.~it remains a destabilizing effect (later on we will see that this is not a strict stability condition).
Yet the magnitude of the destabilizing effect is reduced when $\alpha\!_{_{LD}}\!>\!0$ (i.e.~it is determined by $f-\alpha\!_{_{LD}}$ instead of $f$ alone), which indicates that the modification in Eq. \eqref{eq:rsf_phi1}, with $\alpha\!_{_{LD}}\!>\!0$, promotes stability. Consequently, the simple considerations discussed here suggest that replacing Eqs. \eqref{eq:rsf_phi} and \eqref{eq:rsf_stress} with Eqs. \eqref{eq:gradual_sigma1}-\eqref{eq:rsf_phi1}, for $\alpha\!_{_{LD}}, \alpha\!_{_{PC}}\!>\!0$, will facilitate stability.

Next, we aim at rigorously studying the generalized model,
which incorporates the modified response to normal stress variations in Eqs.~\eqref{eq:gradual_sigma1}-\eqref{eq:rsf_phi1},
in the large $k H$ limit. For that aim, we first derive the linear stability spectrum for this case. Contrary to the analysis in Sect.~\ref{sec:infiniteH}, we should explicitly include now perturbations of the friction stress $\tau(x,t)\!=\!\tau_0+A_\tau\exp\!{[\Lambda t - ikx]}$ since $\tau(x,t)$ satisfies a dynamical equation of its own. The linear stability spectrum reads
\begin{eqnarray}
\mu\,k_d\,g_1\left(1 + \alpha\!_{_{PC}} \Lambda \frac{D}{v_0} \right) - i\,\mu\,k\,f\,g_2  +\frac{i\,k\,\mu\,\alpha\!_{_{LD}}\,g_2\,\Lambda}{\Lambda + \frac{v_0}{D}} + \frac{\sigma_0\,\Lambda}{\Lambda + \frac{v_0}{D}} \left( \Lambda \pa_v\!f + \frac{v_0}{D} d_v\!f \right) = 0 \ ,
\label{eq:spectrum_dim_generalized}
\end{eqnarray}
which reduces to Eq.~\eqref{eq:spectrum_dim} for $\alpha\!_{_{LD}}\!=\!\alpha\!_{_{PC}}\!=\!0$, $|g'(1)|\!=\!1$
and $g_{1,2}\!\to\!G_{1,2}$. The non-dimensional spectrum takes the form
\begin{equation}
\gamma \left(1 -i q z \right) \left[\sqrt{1-\beta^2 z^2}\, g_1(z,\beta)\left(1 -i\,\alpha\!_{_{PC}}\,q\,z\right) - i f g_2(z,\beta) \right]
+\gamma\,q\,\alpha\!_{_{LD}}\,g_2(z,\beta)\,z- i z \left(\Delta -i q z \right) = 0 \ ,
\label{eq:spectrum_dimless_generalized}
\end{equation}
which reduces to Eq. \eqref{eq:spectrum_dimless} when $\alpha\!_{_{LD}}\!=\!\alpha\!_{_{PC}}\!=\!0$. Our next goal would be to analyze this linear stability spectrum.

\subsection{Dilatational wave dominated instability}

To analyze Eq. \eqref{eq:spectrum_dimless_generalized}, we closely follow the procedure described in Sect.~\ref{subsec:dilatational}.
We first look for solutions located near $z\!=\!1/\beta$, corresponding to a dilatational wave dominated instability.
The expansions in Eqs. \eqref{eq:kappa_d} and \eqref{eq:kappad_expansion} remain valid and can be substituted into Eq. \eqref{eq:spectrum_dimless_generalized} to yield
\begin{equation}
\kappa_d \simeq \frac{i f \gamma \left(1-i q/\beta \right)\left(2-1/\beta^2 \right)+ i\left(\Delta - iq/\beta \right)\!/\beta -\gamma\,q \,\alpha\!_{_{LD}}\left(2-1/\beta^2\right)\!/\beta}{\gamma\left(1- i q/\beta \right)\left(1 \mp f \sqrt{1/\beta^2 -1}-i\,\alpha\!_{_{PC}}\,q/\beta \right)\!/\beta^2 \mp i\,\gamma\,q\,\alpha\!_{_{LD}}\sqrt{1/\beta^2-1}/\beta^3} \ .
\label{eq:kappa_d2}
\end{equation}
Here, as before, the stable solution corresponds to the minus sign and unstable one to the plus sign. Equation \eqref{eq:kappa_d2} is then substituted in Eq. \eqref{eq:kappa_d} to obtain $z$, from which the growth rate is derived according to $\Re(\lambda)\!=\!q\Im(z)$.

While the resulting analytic expression for $\Re(\lambda)$ is readily available, it is a bit lengthy and we do not report it
explicitly here. Yet, some analytic insight can be gained directly from Eq.~\eqref{eq:kappa_d2}.
First, we note that $\ald$ always appears through the combination $f-\ald$;
in the numerator it appears through $\gamma\,q \,(f-\ald)(2-1/\beta^2)\!/\beta$ and in the denominator through
$i\,\gamma\,q\,(f-\ald)\sqrt{1/\beta^2-1}/\beta^3$. This is in line with the qualitative discussion above,
which indicated that the main effect of $\ald$ is to reduce the effective friction coefficient. However, it is crucial to understand
that while $\ald$ enters the problem through the combination $f-\ald$, $f$ also appears independently
(cf. the first term in the numerator of Eq.~\eqref{eq:kappa_d2}). Furthermore, while $f-\ald$ is always multiplied
by the wave-number $q$, $f$ appears independently of $q$. This structure will have direct implications for the
stability boundary, which corresponds to $q\!\to\!0$, as will be discussed below.

The other new parameter, $\apc$, is also multiplied by $q$ in Eq.~\eqref{eq:kappa_d2}.
In fact, it introduces a new term proportional to $\apc q^2$ in the denominator of Eq.~\eqref{eq:kappa_d2},
which does not exist in the theory with $\apc\!=\!0$. This has interesting implications for the behavior of the growth
rate as $q\!\to\infty$. Our previous analysis for $\apc\!=\!0$ showed that $\Re(\lambda)$ approaches a finite constant
as $q\!\to\!\infty$. For $\apc\!>\!0$, the presence of the new term proportional to $\apc q^2$ in the denominator
of Eq.~\eqref{eq:kappa_d2} and the fact that the largest power of $q$ in the numerator is linear,
implies that $\Re(\lambda)\!\to\!0$ as $q\!\to\!\infty$. This suggests that even if all wave-numbers are unstable, the growth rate vanishes for sufficiently large $q$. This property of the generalized model is appealing from a basic physics perspective and as such it constitutes an improvement relative to the standard rate-and-state friction model.

\begin{figure}[here]
\centering
\begin{tabular}{ccc}
\includegraphics[width=0.5\textwidth]{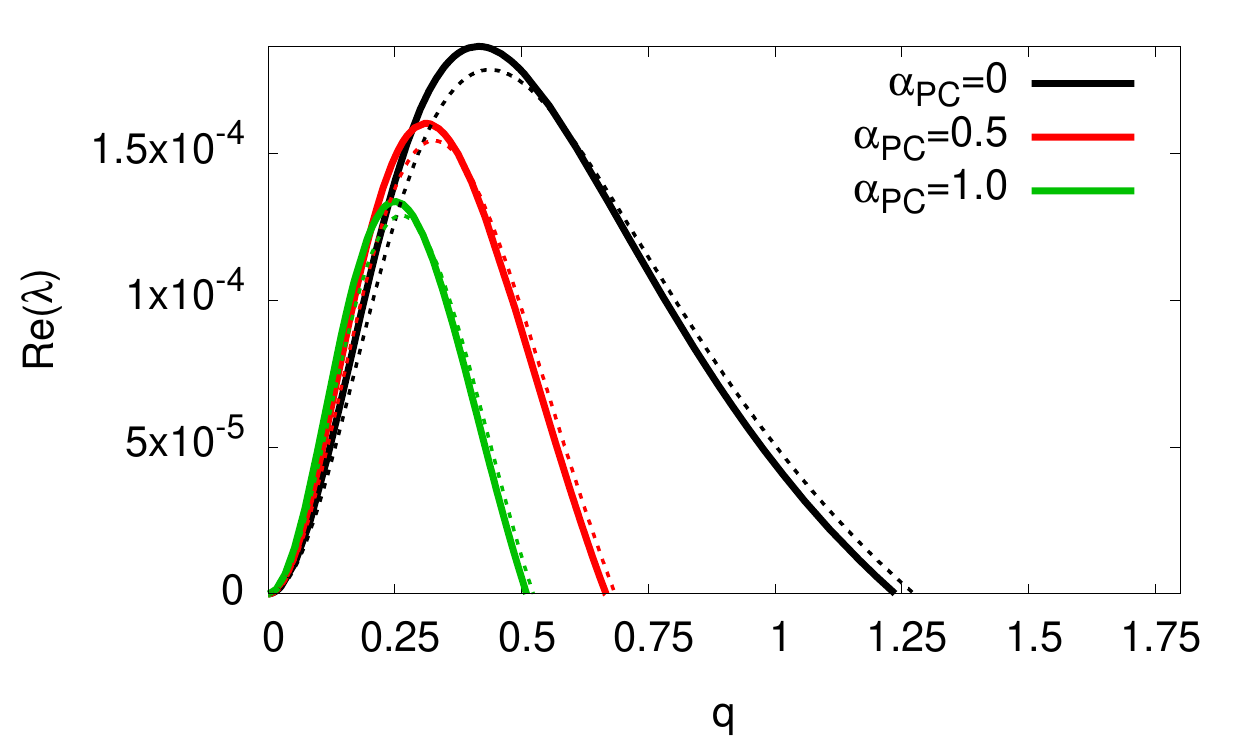}
\includegraphics[width=0.5\textwidth]{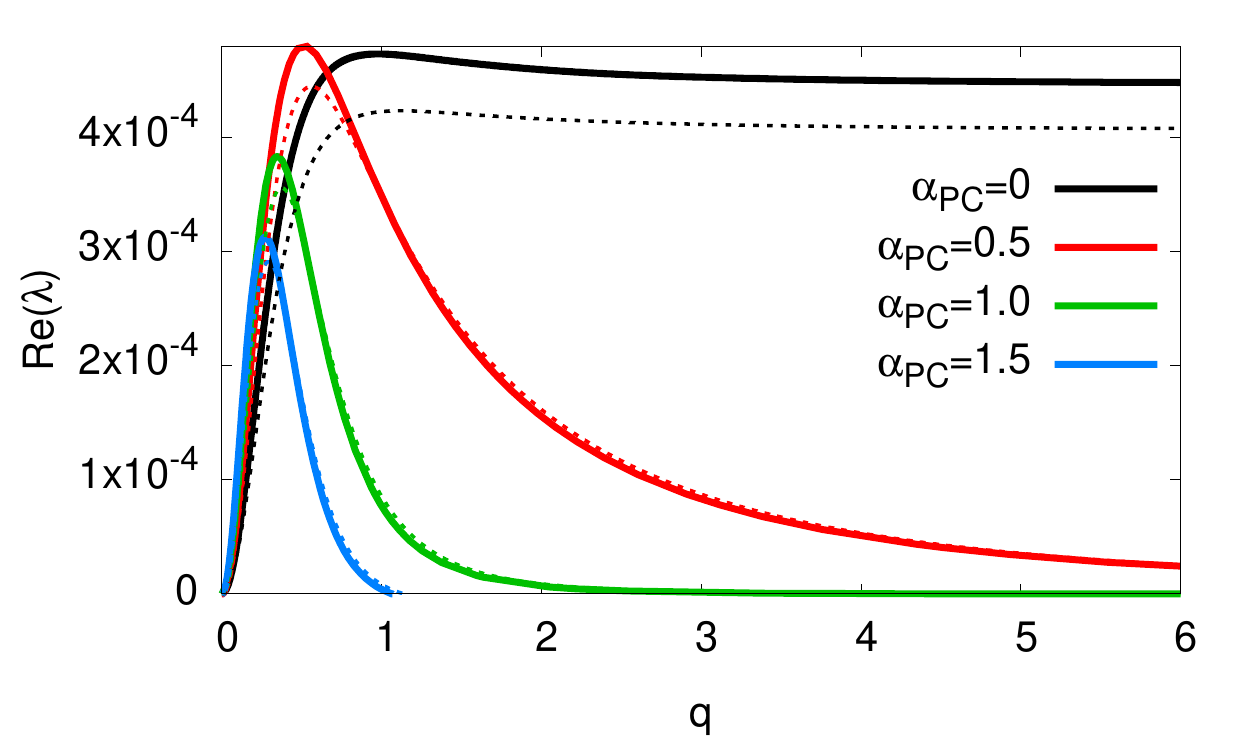}
\end{tabular}
\caption{The dimensionless growth rate $\Re(\lambda)$ vs. wave-number $q$ from a direct numerical solution of Eq.~\eqref{eq:spectrum_dimless_generalized} (thick lines) and the analytic prediction corresponding to Eq.~\eqref{eq:kappa_d} with $\kappa_d$ of Eq.~\eqref{eq:kappa_d2} (thin lines). Here we set $\ald\!=\!0$ and use various values of $\apc$. In the left panel we use $\gamma\!=\!3.2$ and in the right one $\gamma\!=\!4$ such that $\Gamma\!=\!0.96$ and $1.2$, respectively (the rest of the parameters are as in Fig.~\ref{fig:finiteH}, i.e. $f\!=\!0.3,\ \beta\!=\!0.5$ and $\Delta\!=\!0.7$.).}
\label{fig:modified1}
\end{figure}

The analytic properties discussed above are demonstrated in Fig.~\ref{fig:modified1}, where we show $\Re(\lambda)(q)$ as obtained from a direct numerical solution of
Eq.~\eqref{eq:spectrum_dimless_generalized} and the analytic prediction ($\Re(\lambda)\!=\!q\Im(z)$, where $z$ is given in Eq. \eqref{eq:kappa_d} with $\kappa_d$ of Eq. \eqref{eq:kappa_d2}) for two sets of the parameters $(\Delta, f, \beta, \gamma)$, $\ald\!=\!0$ and various values of $\apc$ (only $\Re(\lambda)\!>\!0$ is shown). That is, we isolate the effect of $\apc$ on the stability problem. A few conclusions can be drawn from the figure.
First, we observe that the analytic prediction is in a very good quantitative agreement with the exact numerical solution for all of the parameters considered, yet again lending strong support to the theoretical approach. Second, as expected, we observe that increasing $\apc\!>\!0$ reduces the range of instability and the magnitude of the growth rate (i.e.~it promotes stability). Finally, we observe (right panel) that as predicted analytically, $\Re(\lambda)$ decays to zero at large $q$'s even when all $q$'s are unstable. We also observe that an infinite range of unstable modes can become finite upon increasing $\apc$.

\begin{figure}[here]
\centering
\begin{tabular}{ccc}
\includegraphics[width=0.5\textwidth]{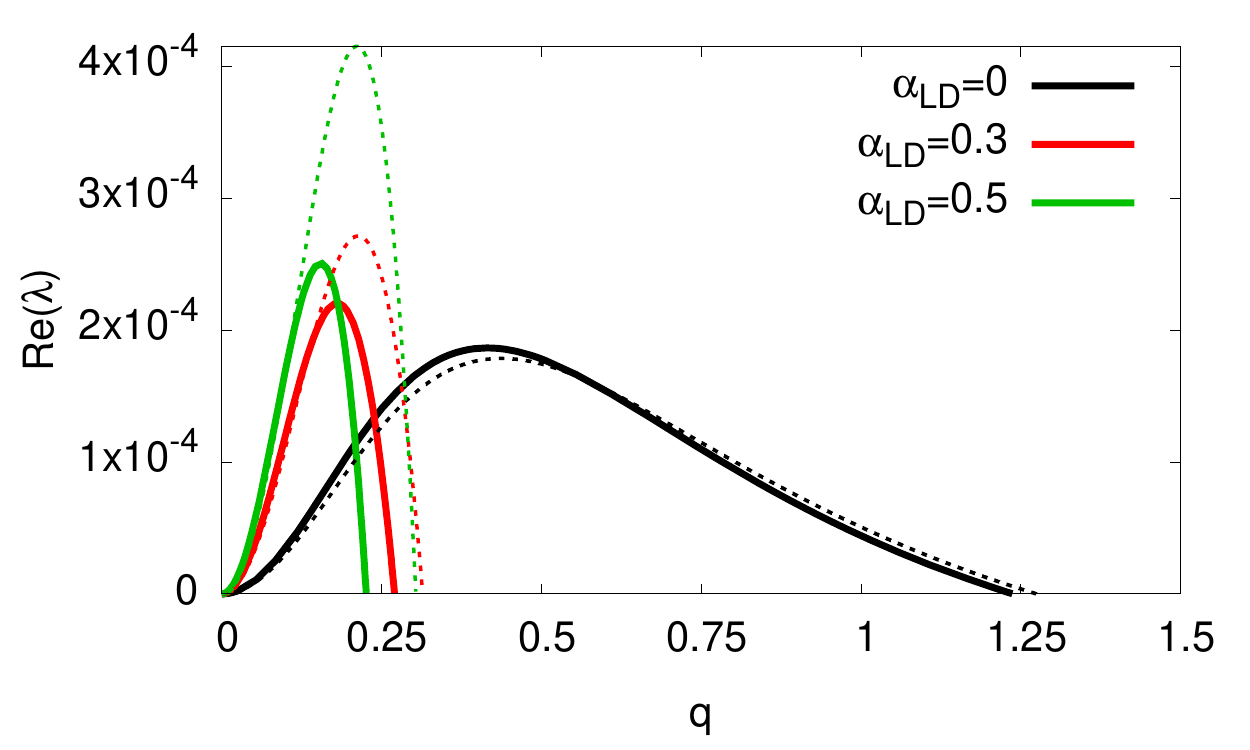}
\includegraphics[width=0.5\textwidth]{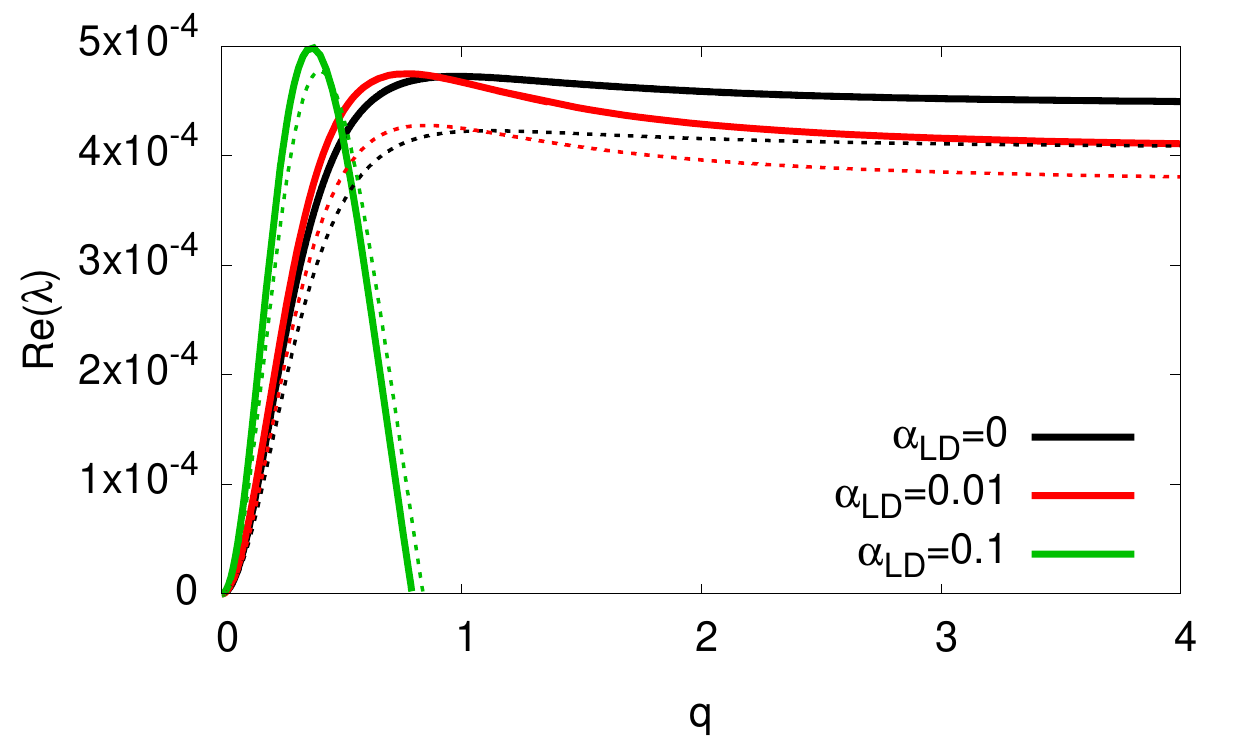}
\end{tabular}
\caption{The same as Fig. \ref{fig:modified1}, but with $\apc\!=\!0$ and varying values of $\ald$.}
\label{fig:modified2}
\end{figure}

In Fig.~\ref{fig:modified2} we elucidate the effect of $\ald$ on the stability problem.
We follow the same scheme as in Fig. \ref{fig:modified1}, but now set $\apc\!=\!0$ and vary $\ald$.
We again observe that the analytic prediction provides a good quantitative approximation to the exact numerical solution (some deviations are observed with increasing $\ald$ in the left panel). We also observe that as the simplified analysis above predicted, increasing $\ald$ promotes stability. Note that since we consider here $\apc\!=\!0$, $\Re(\lambda)$ approaches a constant as $q\!\to\!\infty$ because a finite $\ald$ does not introduce higher order powers of $q$ into the expression for
$\kappa_d$ in Eq.~\eqref{eq:kappa_d2}, as compared to the $\ald\!=\!0$ case discussed in Sect.~\ref{subsec:dilatational}.
As Fig.~\ref{fig:modified2} clearly shows (right panel), the value and sign of the constant does depend on $\ald$.
Finally, we note that an instability exists also when $f-\ald\!<\!0$ (see left panel),
demonstrating that $\ald\!>\!f$ does not eliminate altogether the destabilizing elastodynamic bi-material effect,
as implied by the simplified analysis at the beginning of this section.

The last issue we need to discuss is the threshold condition (stability boundary) associated with the dilatational wave
dominated instability. As we have seen in relation to Eq.~\eqref{eq:kappa_d1}, the threshold condition emerges from the sign of the $q$-independent part of the numerator of $\Im[\kappa_d]$ in the latter equation. On the other hand, as discussed above, $\apc$ and $\ald$ in Eq.~\eqref{eq:kappa_d2} are multiplied by $q$, which implies that they do not affect the threshold condition. Consequently, Eq. \eqref{eq:threshold} remains the threshold condition for the dilatational wave dominated instability in the generalized model, independently of $\apc$ and $\ald$. In the next subsection, we discuss the threshold condition associated with the shear wave dominated instability within the generalized model.

\subsection{Shear wave dominated instability}

We now consider the shear wave dominated instability, which corresponds to solutions near $z\!=\!-1$ in the complex plane. Repeating the analysis that led to Eq.~\eqref{eq:full_shear_inst}, the critical wave-number and the stability threshold read
\begin{align}
q_{_c}^{_{(s)}} = \sqrt{\frac{1}{\apc} \!+\! \left[\!\frac{1\!-\!\gamma(f\!-\!\ald)}{2\,\apc\gamma\sqrt{1-\beta^2}}\!\right]^2} \!-\!  \frac{1\!-\!\gamma(f\!-\!\ald)}{2\,\apc\gamma\sqrt{1-\beta^2}} \qquad\hbox{and}\qquad \Delta \!=\! \gamma f \!-\! \gamma \sqrt{1-\beta^2} (1+\apc)\,q_{_c}^{_{(s)}} \ .\nonumber\\
 \label{eq:full_shear_inst1}
\end{align}
We first observe that the stability threshold, unlike the one for the dilatational wave dominated instability, depends explicitly on both $\apc$ and $\ald$
(the latter only through $q_{_c}^{_{(s)}}$). This happens because the instability occurs at a finite wave-number.
Equation~\eqref{eq:full_shear_inst1} shows that while $q_{_c}^{_{(s)}}$ depends on $f$ and $\ald$ only through the combination $f-\ald$ (in addition to its dependence on $\apc$),
the threshold condition depends both on $f$ and $f-\ald$.

The limit $\apc\!\to\!0$ should be handled with some care as a few terms in Eq.~\eqref{eq:full_shear_inst1} appear to diverge upon a naive substitution of $\apc\!=\!0$.
Nevertheless, this limit exists whenever $1\!-\!\gamma(f-\ald)\!>\!0$ and takes the form
\begin{equation}
q_{_c}^{_{(s)}} = \frac{\gamma\sqrt{1-\beta^2}}{1-\gamma(f-\ald)}  \quad\qquad\hbox{and}\quad\qquad  \Delta = \gamma f - \frac{\gamma^2(1-\beta^2)}{1-\gamma (f-\ald)} \ .
\label{eq:full_shear_inst2}
\end{equation}
Finally, Eq.~\eqref{eq:full_shear_inst} is recovered once $\ald\!=\!0$ is substituted in the above expressions.
With this we complete the large $k H$ analysis of the generalized models in the elastodynamic regime
(we do not consider here the quasi-static limit).

\section{Brief discussion and concluding remarks}
\label{sec:summary}

In this paper we considered in quite general terms the linear stability of homogeneous sliding along frictional interfaces separating strongly dissimilar elastic materials, with a focus on finite size effects, elastodynamic effects and velocity-strengthening friction. The linear stability problem is studied, analytically for the most part, within the constitutive framework of generalized rate-and-state friction models, including velocity-weakening associated with the maturity/age of contact asperities, instantaneous rheological strengthening, steady-state velocity-strengthening and a regularized response to normal stress variations. We considered finite size systems, of height $H$, and analyzed the stability spectrum in both the small and large $k H$ limits. The various competing physical effects, most notably the destabilizing elastodynamic bi-material effect and stabilizing effects associated with the interfacial constitutive behavior, are quantified through several dimensionless parameters.

We showed that there exists a universal instability (independent of the details of the friction law, but possibly dependent on the assumed large material contrast) characterized by a wave-number $k\!\sim\!H^{-1}$ and a maximal growth rate $\Re[\Lambda]\!\propto\!f c_s/H$, mediated by waveguide-like modes. The role of boundary conditions for finite size systems has been also highlighted. For large systems, $H\!\to\!\infty$, we provided a comprehensive quantitative picture of the stability phase diagram. We showed that in addition to the previously derived quasi-static instability modes ~\citep{Rice2001}, which exist at sufficiently small sliding velocities, there exist also dilatational wave dominated instability modes propagating in the opposite direction to the sliding direction and shear wave dominated instability modes propagating in the sliding direction. The former appear to be the dominant instability modes over a broad range of physical parameters. Our results allow to determine, for every set of physical parameters relevant to a specific frictional system, which of the discussed instabilities features the largest growth rate. In a certain parameter range the instability is manifested through unstable modes at all wave-numbers, yet the frictional response is shown to be mathematically well-posed. The stabilizing roles played by a regularized response to normal stress variations are quantitatively accounted for.

All in all, our analysis shows that steady sliding along strong bi-material frictional interfaces is quite generically unstable, even in the presence of steady-state velocity-strengthening friction of a general form. The corresponding problem for steady-state velocity-weakening friction is known to be generically unstable, but this instability is of different origin compared to the instabilities discussed in this paper. The discussed instabilities depend crucially on the destabilizing bi-material effect and consequently these generic instabilities are not expected to persist in frictional interfaces separating identical materials. Our results should be relevant to any frictional system exhibiting strong material contrast along the frictional interface (and possibly also to finite contrast interfaces) and steady-state velocity-strengthening, for example geophysical systems such as mature earthquake faults~\citep{Ben-Zion2001, Ben-Zion2008} and engineering/tribological systems such as an elastic brake pad sliding on a rigid substrate~\citep{Behrendt2011}.

The results of the linear stability analysis are possibly related to rupture dynamics along bi-material interfaces~\citep{Ben-Zion2001, Ben-Zion2008}. While the exact relations can be elucidated by following the instabilities into the nonlinear regime, most probably only numerically, one can {\em speculate} about these relations. For example, the dilatational wave dominated instability, which propagates in the direction opposite to the sliding motion and emerges at $k\!\to 0$ (i.e. large wavelength), may correspond to crack-like (non-localized) super-shear rupture fronts propagating in the so-called ``un-preferred direction'' (i.e. opposite to the slip direction of the more compliant material, for any material contrast). Likewise, the shear wave dominated instability, which propagates in the direction of sliding motion and emerges at a finite $k$, may correspond to pulse-like (localized) shear rupture fronts propagating in the so-called ``preferred direction'' (i.e. the slip direction of the more compliant material, for any material contrast). These possible relations can be tested experimentally.

In the future, it would be interesting to quantify how the existence of a finite bi-material contrast affects the stability analysis presented here for the large contrast limit, where the destabilizing elastodynamic bi-material effect is the largest. Furthermore, other theoretical issues should be considered; for example, as the dilatational and shear wave dominated instabilities propagate with high velocities in the lab frame of reference, issues of absolute vs. convective instabilities~\citep{LandauPhysKin, Huerre1990} may become important. Moreover, in the presence of lateral boundaries (i.e. for realistic systems of finite size $L$ in the $x$-direction) the so-called global instability~\citep{LandauPhysKin, Huerre1990} may be relevant due to the lateral boundary conditions which induce a coupling between different linear propagating modes. Finally, the finite size $L$ naturally implies a modification of some of the small wave-number instabilities discussed above.

\vspace{0.5cm}

\noindent{\bf Acknowledgements}\\

E.B. acknowledges support of the James S. McDonnell Foundation, the Minerva Foundation with funding from the Federal German Ministry for Education and Research and the William Z. and Eda Bess Novick Young Scientist Fund. This research is made possible in part by the historic generosity of the Harold Perlman Family. R.S. acknowledges support by the DFG priority program 1713.

\vspace{0.5cm}

\appendix
\numberwithin{equation}{section}

\noindent{\bf Appendix}

\section{Derivation of Eq.~\eqref{eq:Hspectrum_expansion}}
\label{appendix_derivation}

We start by expanding first the basic functions appearing in Eq.~\eqref{eq:Hspectrum_dimensionless}
(see also Eq.~\eqref{eq:pert_elastodynamics_auxiliary}) as
\begin{equation}
\begin{split}
 k_d(\Lambda,k)&\simeq\pm \frac{\Lambda_0}{c_d}  \ , \qquad
 k_s(\Lambda,k)\simeq\pm \frac{\Lambda_0}{c_s}  \ ,\\
 \coth(H k_d)&\simeq\pm \frac{H \delta\Lambda}{c_d} \ , \qquad
 \tanh(Hk_s)\simeq\pm \tanh(H\Lambda_0/c_s) \ ,
\end{split}
\label{eq:expansion_intermediate}
\end{equation}
where the $\pm$ correspond to the stable/unstable branches, respectively. As will be seen right away,
the distinction between the two branches disappears in the final expression.
Using these expansions, we obtain the following relevant combinations
\begin{equation}
\begin{split}
 k_s(\Lambda,k)\,\tanh(H\,k_s)&\simeq \frac{\Lambda_0}{c_s}\tanh(H\Lambda_0/c_s)  \ ,\\
 k_s(\Lambda,k) k_d(\Lambda,k) &\simeq \frac{\Lambda_0^2}{c_s\,c_d}  \ ,\\
\coth(H k_d) \tanh(H k_s)&\simeq\frac{H \delta\Lambda}{c_d} \tanh(H\Lambda_0/c_s)   \ .
\end{split}
\end{equation}
These expressions allow us to expand the three contributions to $S(\Lambda,k, H)$ (after dropping the common prefactor $\mu$).

First, we consider the contribution to $S(\Lambda, k, H)$ in Eq.~\eqref{eq:Hspectrum_dimensionless}, associated with $\delta\sigma_{yy}$,
\begin{align}
i\,k\,f\,&G_2(\Lambda,k, H) = - i\,k\,f \left(2- \frac{G_1(\Lambda, k, H)}{\coth(Hk_d)} \right) = - i\,k\,f \left(2- \frac{\Lambda ^2/c_s^2 }{k_d k_s \coth(Hk_d)\tanh(Hk_s)-k^2} \right)\nonumber\\
&\simeq  i\,k\,f \frac{\Lambda^2/c_s^2 }{k_d k_s \coth(Hk_d)\tanh(Hk_s)}\!\simeq\! \frac{i\,f\,c_s\,k}{\beta^2 H \tanh(H\Lambda_0/c_s) \delta\Lambda}= \pm \frac{f\,c_s\,k}{\beta^2 H \tan(H|\Lambda_0|/c_s) \delta\Lambda} \ .
\end{align}
It is important to note that the latter depends on the {\em ratio} of the two small quantities $k$ and $\delta\Lambda$ and hence contributes to zeroth order. Therefore, in what follows we can expand quantities to zeroth order.

For the contribution associated with $\delta\sigma_{xy}$, we have
\begin{align}
k_d(\Lambda,k)\,G_1(\Lambda,k, H) \!=\! \frac{k_d\coth(Hk_d)\Lambda^2/c_s^2 }{k_d k_s \coth(Hk_d)\tanh(Hk_s)-k^2} \!\simeq\!
\frac{\Lambda ^2/c_s^2 }{k_s\,\tanh(Hk_s)}\!\simeq\! \frac{|\Lambda_0|}{c_s\,\tan(H|\Lambda_0|/c_s)},\nonumber\\
\end{align}
where in the last step we used the fact that $\Lambda_0$ is purely imaginary, which implies that
$\tanh(H\Lambda_0/c_s)\!=\!\pm\, i \tan(H|\Lambda_0|/c_s)$ ($\pm$ here correspond to $\Lambda_0\!=\!\pm\, i |\Lambda_0|$). We thus conclude that the contribution corresponding
to $\delta\sigma_{xy}$ is finite and independent of the sign of $\Im(\Lambda_0)$. Note that we started by expanding around
$(\Lambda\!=\!\Lambda_0,k\!=\!0)$, which corresponds to $\delta\sigma_{xy}(\Lambda, k\!\to\!0)\!=\!0$ (see Eq.~\eqref{eq:waveguide_cutoff} and the discussion around it), and now we find that $\delta\sigma_{xy}(\Lambda\!\to\!\Lambda_0, k\!\to\!0)$ is actually finite. There is, however, no contradiction here.
The point is that previously we used $\delta\sigma_{xy}(\Lambda, k)\!=\!0$ to {\em derive} the dispersion relation
$\Lambda_{wg}(k)$ and then took the $k\!\to\!0$ limit, and here we treat $\delta\sigma_{xy}(\Lambda, k)$ as
a function of two independent variables, where $\Lambda(k)$ is not known (the goal is to determine it from $S(\Lambda,k,H)\!=\!0$).

 Finally, the friction law contribution is simply evaluated
at $\Lambda\!=\!\Lambda_0$, i.e.
\begin{eqnarray}
\frac{\Lambda}{c_s}\frac{\lambda(\Lambda)+\Delta}{\gamma\big(\lambda(\Lambda)+1\big)}
\simeq \frac{\Lambda_0}{c_s}\frac{\lambda_0+\Delta}{\gamma\big(\lambda_0+1\big)}
=\frac{|\Lambda_0|}{c_s \gamma}\left(\frac{-|\lambda_0|(1-\Delta)\pm i (\Delta +|\lambda_0|^2)}{1+|\lambda_0|^2}\right) \ ,
\label{eq:expansion_friction}
\end{eqnarray}
where $\lambda_0\!\equiv\!\tfrac{D\Lambda_0}{|g'(1)|v_0}$ and again the $\pm$ correspond to $\Lambda_0\!=\!\pm\, i |\Lambda_0|$. Collecting all three contributions we end up with Eq.~\eqref{eq:Hspectrum_expansion} of the main text.

\section{Finite height analysis with imposed shear stress}
\label{appendix_BC}

\begin{figure}[here]
\centering
\begin{tabular}{ccc}
\includegraphics[width=0.6\textwidth]{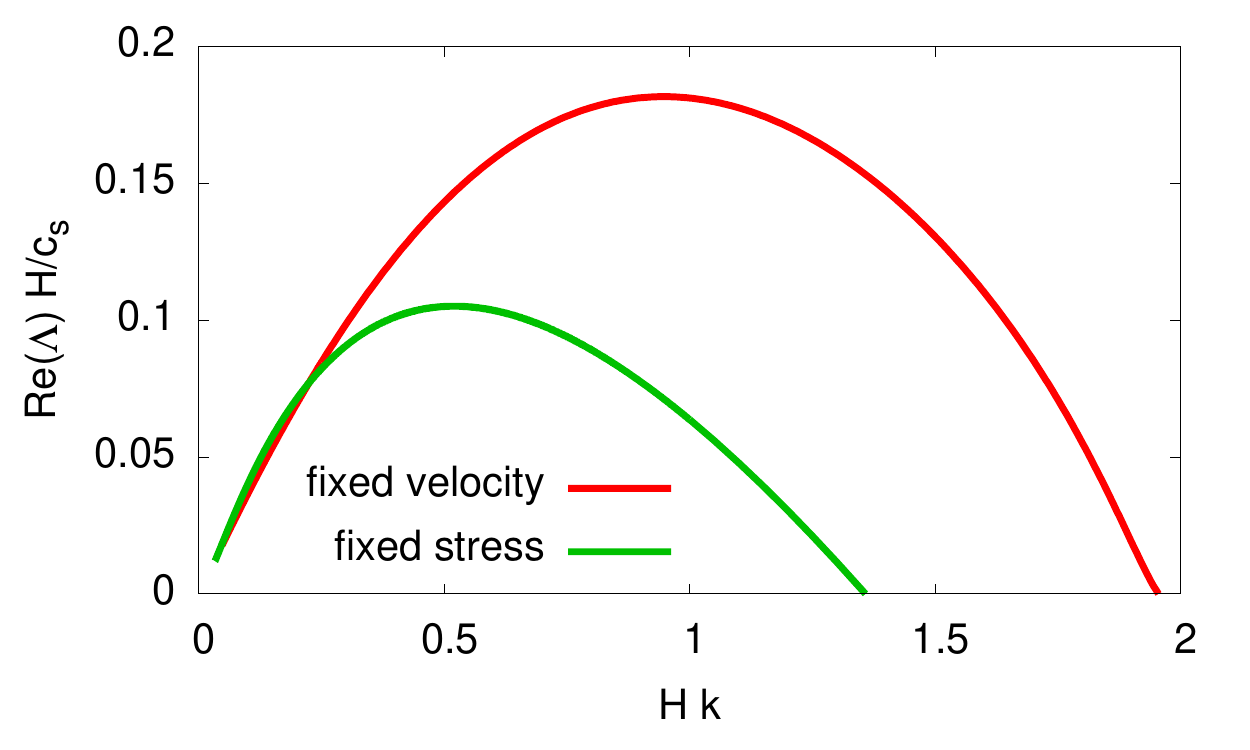}
\end{tabular}
\caption{The growth rate $\Re[\Lambda]$ (in units of $c_s/H$) vs. $H k$, for small $H k$, as in the left panel of Fig.~\ref{fig:finiteH}. The upper (red) curve corresponds to velocity-controlled boundary conditions (it already appeared in the left panel of Fig.~\ref{fig:finiteH}, with $\gamma f\!=\!0.9$ and $n\!=\!0$, though the range of $H k$ is smaller here). The lower (green) curve corresponds to stress-controlled boundary conditions (see text for details).}
\label{fig:stress-controlled}
\end{figure}

The analysis of finite size systems (i.e.~in small $k H$ limit) in Sect.~\ref{sec:finiteH} was performed under an imposed tangential velocity boundary condition at $y\!=\!H$. As mentioned above, we expect the type of boundary conditions at $y\!=\!H$ to affect the stability problem. To demonstrate this, we consider two situations which are identically the same (in terms of both parameters and geometry) except that in one case we impose a tangential velocity $v_0$ at $y\!=\!H$ (i.e.~as on the left of Eq.~\eqref{eq:2BC}, which was adopted throughout the manuscript until now) and in the other case we impose a shear stress $\tau_0$ at $y\!=\!H$ (i.e.~as on the right of Eq.~\eqref{eq:2BC}). Here $v_0$ and $\tau_0$ are related by the friction law under steady-state conditions. The spectrum equation for the imposed shear stress case is derived along the same lines as for the impose tangential velocity, but the result is a bit too lengthy to be reported on explicitly here. Furthermore, we restrict ourselves in this Appendix to its numerical analysis, though in principle analytic progress in the spirit of the analysis performed in the manuscript can be pursued. In Fig.~\ref{fig:stress-controlled} we plot the normalized growth rate $H\Re[\Lambda]/c_s$ as a function of the normalized wave-number $k H$ for the two cases. It is evident that the result quantitatively depends on the type of boundary condition, though no marked qualitative differences are observed.

\section{Discontinuities (gaps) in the spectrum: Non-localized modes}
\label{appendix_gap}

The full numerical solution of the implicit linear stability spectrum in the $k H\!\to\!\infty$ limit, shown on Fig.~\ref{fig:comparison}, exhibits a discontinuity (gap) at the transition from the unstable to the stable part of the solution. Our goal here is to propose an explanation for the origin of the gap in the spectrum. As a background, note that
while in Sect.~\ref{sec:infiniteH} the $H\!\to\!\infty$ limit was taken directly at the level of the linear stability spectrum of Eq.~\eqref{eq:Hspectrum_dimensionless}, cf. Eq.~\eqref{eq:g12}, one could alternatively take the limit at the level of the elastodynamic solutions of Eqs.~\eqref{eq:elastic fields}. In this case, we demand that the perturbations decay away from the interface, i.e.~as $y\!\to\!\infty$, which implies that $A_3\!=\!A_4\!=\!0$ and only $A_{1,2}$ remain finite. Imposing the two boundary conditions at the interface, $y\!=\!0$, yields the instability spectrum in Eq.~\eqref{eq:spectrum_dimless}.

In this $H\!\to\!\infty$ limit, only the decaying solutions $\{\exp[-k_s y], \exp[-k_d y]\}$ are physically relevant. We will now show that the gap in the spectrum is related to an {\em increasing} solution, which is non-physical in the strict $H\!\to\!\infty$ limit. In particular, we consider a solution involving $\{\exp[k_s y], \exp[-k_d y]\}$, that is
\begin{equation}
\begin{pmatrix}
 \delta u_x(x,y,t)\\ \delta u_y(x,y,t)
 \end{pmatrix}
 =
 \begin{pmatrix}
 k & k_s \\
 - ik_d & ik
 \end{pmatrix}
  \begin{pmatrix}
   A_2 \exp[-k_d y] \\
   A_3 \exp[k_s y]
  \end{pmatrix}
  \exp\!{[\Lambda t - ikx]} \ .
\label{eq:elastic fields_gaps}
\end{equation}

By imposing the boundary conditions at $y\!=\!0$ and properly nondimensionalizing all relevant quantities we obtain the following linear stability spectrum equation
\begin{eqnarray}
\tilde{s}(z, q) \equiv \gamma \left(1 -i q z \right) \left[-\sqrt{1-\beta^2 z^2}\,\tilde{g}_1(z, \beta) - i f \tilde{g}_2(z, \beta) \right]- i z \left(\Delta -i q z \right) = 0 \ ,
\label{eq:spectrum_dimless_gap}
\end{eqnarray}
where
\begin{eqnarray}
\tilde{g}_1(z, \beta) = -\frac{z^2}{1+\sqrt{1-z^2} \sqrt{1-\beta^2 z^2}} \quad\qquad\hbox{and}\quad\qquad \tilde{g}_2(z, \beta) = 2+\tilde{g}_1(z, \beta) \ .
\label{eq:g12_dimless_gap}
\end{eqnarray}
which are the counterparts of Eqs.~\eqref{eq:spectrum_dimless}-\eqref{eq:g12_dimless}.
\begin{figure}[here]
\centering
\begin{tabular}{ccc}
\includegraphics[width=0.6\textwidth]{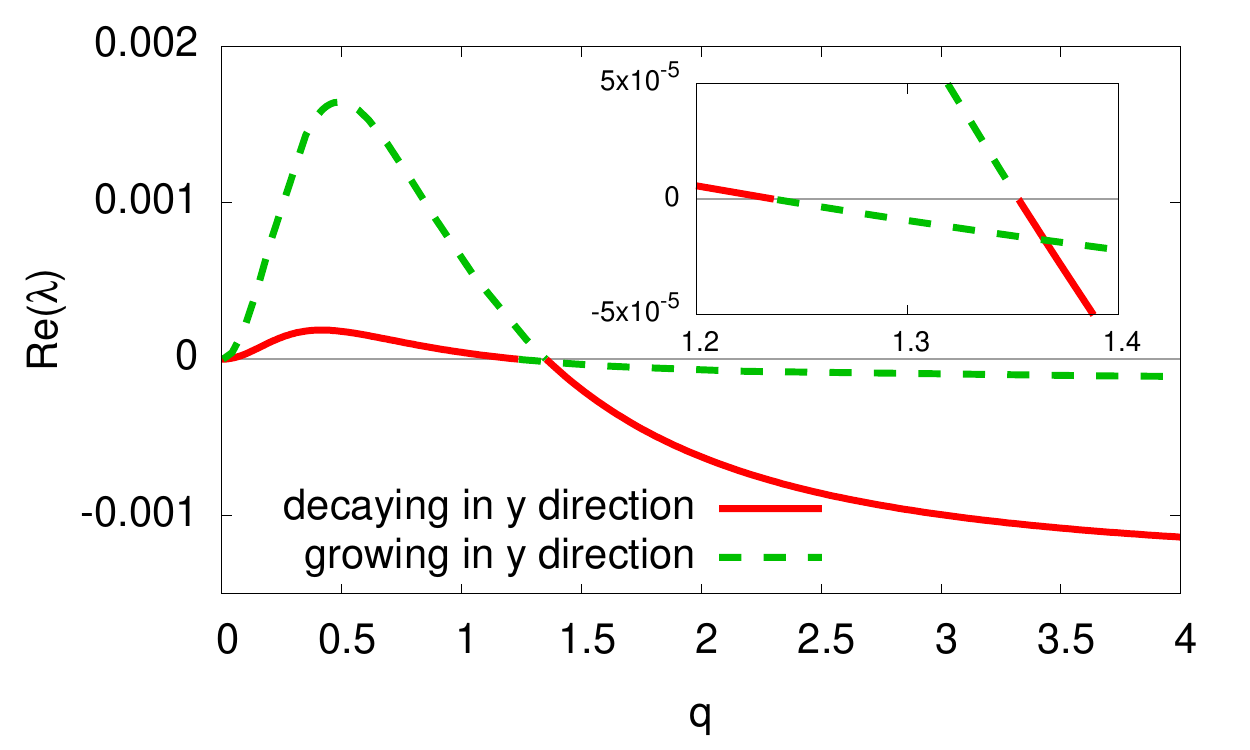}
\end{tabular}
\caption{The dimensionless growth rate $\Re(\lambda)\!=\!q \Im(z)$ vs. the dimensionless wave-number $q$, as in the left panel of Fig.~\ref{fig:comparison}. The solid (red) line is identical to the $\gamma f\!=\!0.96$ curve in the left panel of Fig.~\ref{fig:comparison}. The dashed (green) curve corresponds to solution of the spectrum in Eq.~\eqref{eq:spectrum_dimless_gap} for the same set of parameters.}
\label{fig:gaps}
\end{figure}

In Fig.~\ref{fig:gaps} we present numerical solutions of Eqs.~\eqref{eq:spectrum_dimless} and \eqref{eq:spectrum_dimless_gap} for the very same set of parameters. It is observed that while each of these solutions is discontinuous in itself, superposing the two solutions generates two {\em continuous} functions, each of which is composed of two segments from different solutions. We thus propose that the gap observed in Fig.~\ref{fig:comparison} is related to the solution in Eq.~\eqref{eq:elastic fields_gaps}, which features a non-physical exponential divergence as $y\!\to\!\infty$. In a real system, with a finite height $H$ (however large), increasing solutions always exist and will naturally lead to the regularization of the discontinuities that emerge in the strict $H\!\to\!\infty$ limit.


\begin{thebibliography}{76}
\expandafter\ifx\csname natexlab\endcsname\relax\def\natexlab#1{#1}\fi
\expandafter\ifx\csname url\endcsname\relax
  \def\url#1{\texttt{#1}}\fi
\expandafter\ifx\csname urlprefix\endcsname\relax\def\urlprefix{URL }\fi

\bibitem[{Achenbach and Epstein(1967)}]{Achenbach1967}
Achenbach, J., Epstein, H., 1967. {Dynamic interaction of a layer and a
  half-space}. Journal of the Engineering Mechanics Division 93, 27--42.

\bibitem[{Adams(1995)}]{Adams1995}
Adams, G.~G., 1995. {Self-excited oscillations of two elastic half-spaces
  sliding with a constant coefficient of friction}. Journal of Applied
  Mechanics 62~(4), 867.

\bibitem[{Adams(1998)}]{Adams1998}
Adams, G.~G., 1998. {Steady sliding of two elastic half-spaces with friction
  reduction due to interface stick-slip}. Journal of Applied Mechanics 65~(2),
  470.

\bibitem[{Adams(2000)}]{Adams2000}
Adams, G.~G., 2000. {An intersonic slip pulse at a frictional interface between
  dissimilar materials}. Journal of Applied Mechanics 68~(1), 81.

\bibitem[{Adda-Bedia and {Ben Amar}(2003)}]{Adda-Bedia2003}
Adda-Bedia, M., {Ben Amar}, M., 2003. {Self-sustained slip pulses of finite
  size between dissimilar materials}. Journal of the Mechanics and Physics of
  Solids 51~(10), 1849--1861.

\bibitem[{Ampuero and Ben-Zion(2008)}]{Ampuero2008a}
Ampuero, J.-P., Ben-Zion, Y., 2008. {Cracks, pulses and macroscopic asymmetry
  of dynamic rupture on a bimaterial interface with velocity-weakening
  friction}. Geophysical Journal International 173~(2), 674--692.

\bibitem[{Andrews and Ben-Zion(1997)}]{Andrews1997}
Andrews, D., Ben-Zion, Y., 1997. {Wrinkle-like slip pulse on a fault between
  different materials}. Journal of Geophysical Research 102~(B1), 553.

\bibitem[{Baillet et~al.(2005)Baillet, Linck, D’Errico, Laulagnet, and
  Berthier}]{Baillet2005}
Baillet, L., Linck, V., D'Errico, S., Laulagnet, B., Berthier, Y., 2005.
  {Finite element simulation of dynamic instabilities in frictional sliding
  contact}. Journal of Tribology 127~(3), 652.

\bibitem[{{Bar Sinai} et~al.(2012){Bar Sinai}, Brener, and
  Bouchbinder}]{BarSinai2012}
{Bar Sinai}, Y., Brener, E.~A., Bouchbinder, E., 2012. {Slow rupture of
  frictional interfaces}. Geophysical Research Letters 39~(3), L03308.

\bibitem[{Bar-Sinai et~al.(2013)Bar-Sinai, Spatschek, Brener, and
  Bouchbinder}]{Bar-Sinai2013pre}
Bar-Sinai, Y., Spatschek, R., Brener, E.~A., Bouchbinder, E., 2013.
  {Instabilities at frictional interfaces: Creep patches, nucleation, and
  rupture fronts}. Physical Review E 88~(6), 060403(R).

\bibitem[{Bar-Sinai et~al.(2014)Bar-Sinai, Spatschek, Brener, and
  Bouchbinder}]{Bar-Sinai2014jgr}
Bar-Sinai, Y., Spatschek, R., Brener, E.~A., Bouchbinder, E., 2014. {On the
  velocity-strengthening behavior of dry friction}. Journal of Geophysical
  Research: Solid Earth 119~(3), 1738--1748.

\bibitem[{Bar-Sinai et~al.(2015)Bar-Sinai, Spatschek, Brener, and
  Bouchbinder}]{Bar-Sinai2015SciRep}
Bar-Sinai, Y., Spatschek, R., Brener, E.~A., Bouchbinder, E., 2015.
  {Velocity-strengthening friction significantly affects interfacial dynamics,
  strength and dissipation.} Scientific reports 5, 7841.

\bibitem[{Baumberger and Berthoud(1999)}]{Baumberger1999}
Baumberger, T., Berthoud, P., 1999. {Physical analysis of the state- and
  rate-dependent friction law. II. Dynamic friction}. Physical Review B 60~(6),
  3928--3939.

\bibitem[{Baumberger and Caroli(2006)}]{Baumberger2006}
Baumberger, T., Caroli, C., 2006. {Solid friction from stick–slip down to
  pinning and aging}. Advances in Physics 55~(3-4), 279--348.

\bibitem[{Behrendt et~al.(2011)Behrendt, Weiss, and Hoffmann}]{Behrendt2011}
Behrendt, J., Weiss, C., Hoffmann, N.~P., 2011. {A numerical study on
  stick-slip motion of a brake pad in steady sliding}. Journal of Sound and
  Vibration 330~(4), 636--651.

\bibitem[{Ben-David et~al.(2010)Ben-David, Rubinstein, and
  Fineberg}]{Ben-David2010-ageing}
Ben-David, O., Rubinstein, S.~M., Fineberg, J., 2010. {Slip-stick and the
  evolution of frictional strength.} Nature 463~(7277), 76--9.

\bibitem[{Ben-Zion(2001)}]{Ben-Zion2001}
Ben-Zion, Y., 2001. {Dynamic ruptures in recent models of earthquake faults}.
  Journal of the Mechanics and Physics of Solids 49~(9), 2209--2244.

\bibitem[{Ben-Zion(2008)}]{Ben-Zion2008}
Ben-Zion, Y., 2008. {Collective behavior of earthquakes and faults:
  Continuum-discrete transitions, progressive evolutionary changes, and
  different dynamic regimes}. Reviews of Geophysics 46~(4), RG4006.

\bibitem[{Ben-Zion and Andrews(1998)}]{Ben-Zion1998}
Ben-Zion, Y., Andrews, D., 1998. {Properties and implications of dynamic
  rupture along a material interface}. Bulletin of the Seismological Society of
  America 88~(4), 1085--1094.

\bibitem[{Ben-Zion and Huang(2002)}]{Ben-Zion2002}
Ben-Zion, Y., Huang, Y., 2002. {Dynamic rupture on an interface between a
  compliant fault zone layer and a stiffer surrounding solid}. Journal of
  Geophysical Research 107~(B2), 2042.

\bibitem[{Berthoud et~al.(1999)Berthoud, Baumberger, G’Sell, and
  Hiver}]{Berthoud1999}
Berthoud, P., Baumberger, T., G’Sell, C., Hiver, J.-M., 1999. {Physical
  analysis of the state- and rate-dependent friction law: Static friction}.
  Physical Review B 59~(22), 14313--14327.

\bibitem[{Bureau et~al.(2000)Bureau, Baumberger, and Caroli}]{Bureau2000}
Bureau, L., Baumberger, T., Caroli, C., 2000. {Shear response of a frictional
  interface to a normal load modulation}. Physical Review E 62~(5), 6810--6820.

\bibitem[{Cochard and Rice(2000)}]{Cochard2000}
Cochard, A., Rice, J.~R., 2000. {Fault rupture between dissimilar materials:
  Ill-posedness, regularization, and slip-pulse response}. Journal of
  Geophysical Research 105~(B11), 25891.

\bibitem[{Comninou(1977{\natexlab{a}})}]{Comninou1977a}
Comninou, M., 1977{\natexlab{a}}. {Interface crack with friction in the contact
  zone}. Journal of Applied Mechanics 44~(4), 780.

\bibitem[{Comninou(1977{\natexlab{b}})}]{Comninou1977b}
Comninou, M., 1977{\natexlab{b}}. {The interface crack}. Journal of Applied
  Mechanics 44~(4), 631.

\bibitem[{Comninou and Schmueser(1979)}]{Comninou1979}
Comninou, M., Schmueser, D., 1979. {The interface crack in a combined
  tension-compression and shear field}. Journal of Applied Mechanics 46~(2),
  345.

\bibitem[{Crupi and Bizzarri(2013)}]{Crupi2013}
Crupi, P., Bizzarri, A., 2013. {The role of radiation damping in the modeling
  of repeated earthquake events}. Annals of Geophysics 56~(1), R0111.

\bibitem[{{Di Bartolomeo} et~al.(2010){Di Bartolomeo}, Meziane, Massi, Baillet,
  and Fregolent}]{DiBartolomeo2010}
{Di Bartolomeo}, M., Meziane, A., Massi, F., Baillet, L., Fregolent, A., 2010.
  {Dynamic rupture at a frictional interface between dissimilar materials with
  asperities}. Tribology International 43~(9), 1620--1630.

\bibitem[{Dieterich(1978)}]{Dieterich1978}
Dieterich, J.~H., 1978. {Time-dependent friction and the mechanics of
  stick-slip}. Pure and Applied Geophysics 116~(4-5), 790--806.

\bibitem[{Dieterich(1979)}]{Dieterich1979}
Dieterich, J.~H., 1979. {Modeling of rock Friction 1. experimental results and
  constitutive equations}. Journal of Geophysical Research 84~(B5), 2161--2168.

\bibitem[{Dieterich(1992)}]{Dieterich1992}
Dieterich, J.~H., 1992. {Earthquake nucleation on faults with rate-and
  state-dependent strength}. Tectonophysics 211~(1-4), 115--134.

\bibitem[{Dieterich and Kilgore(1994)}]{Dieterich1994}
Dieterich, J.~H., Kilgore, B.~D., 1994. {Direct observation of frictional
  contacts: New insights for state-dependent properties}. Pure and Applied
  Geophysics 143~(1-3), 283--302.

\bibitem[{Dieterich and Linker(1992)}]{Dieterich1992a}
Dieterich, J.~H., Linker, M.~F., 1992. {Fault stability under conditions of
  variable normal stress}. Geophysical Research Letters 19~(16), 1691--1694.

\bibitem[{Gerde and Marder(2001)}]{Gerde2001}
Gerde, E., Marder, M., 2001. {Friction and fracture.} Nature 413~(6853),
  285--288.

\bibitem[{Harris and Day(1997)}]{Harris1997}
Harris, R.~A., Day, S.~M., 1997. {Effects of a low-velocity zone on a dynamic
  rupture}. Bulletin of the Seismological Society of America 87~(5),
  1267--1280.

\bibitem[{Hawthorne and Rubin(2013)}]{Hawthorne2013}
Hawthorne, J.~C., Rubin, A.~M., 2013. {Tidal modulation and back-propagating
  fronts in slow slip events simulated with a velocity-weakening to
  velocity-strengthening friction law}. Journal of Geophysical Research: Solid
  Earth 118~(3), 1216--1239.

\bibitem[{Heslot et~al.(1994)Heslot, Baumberger, Perrin, Caroli, and
  Caroli}]{Heslot1994}
Heslot, F., Baumberger, T., Perrin, B., Caroli, B., Caroli, C., 1994. {Creep,
  stick-slip, and dry-friction dynamics: Experiments and a heuristic model}.
  Physical Review E 49~(6), 4973--4988.

\bibitem[{Huerre and Monkewitz(1990)}]{Huerre1990}
Huerre, P., Monkewitz, P.~A., 1990. {Local and global instability in spatially
  developing flows}. Ann. Rev. Fluid Mech. 22, 473--537.

\bibitem[{Ibrahim(1994{\natexlab{a}})}]{Ibrahim1994a}
Ibrahim, R.~A., 1994{\natexlab{a}}. {Friction-Induced Vibration, Chatter,
  Squeal, and Chaos. Part I: Mechanics of Contact and Friction}. Applied
  Mechanics Reviews 47~(7), 209--226.

\bibitem[{Ibrahim(1994{\natexlab{b}})}]{Ibrahim1994}
Ibrahim, R.~A., 1994{\natexlab{b}}. {Friction-Induced Vibration, Chatter,
  Squeal, and Chaos. Part II: Dynamics and Modeling}. Applied Mechanics
  Reviews 47~(7), 227.

\bibitem[{Ikari et~al.(2013)Ikari, Marone, Saffer, and Kopf}]{Ikari2013}
Ikari, M.~J., Marone, C.~J., Saffer, D.~M., Kopf, A.~J., 2013. {Slip weakening
  as a mechanism for slow earthquakes}. Nature Geoscience 6~(7), 468--472.

\bibitem[{Ikari et~al.(2009)Ikari, Saffer, and Marone}]{Ikari2009}
Ikari, M.~J., Saffer, D.~M., Marone, C.~J., 2009. {Frictional and hydrologic
  properties of clay-rich fault gouge}. Journal of Geophysical Research
  114~(B5), B05409.

\bibitem[{Kato(2003)}]{Kato2003}
Kato, N., 2003. {A possible model for large preseismic slip on a deeper
  extension of a seismic rupture plane}. Earth and Planetary Science Letters
  216~(1-2), 17--25.

\bibitem[{Lifshitz and Pitaevskii(1981)}]{LandauPhysKin}
Lifshitz, E.~M., Pitaevskii, L.~P., 1981. {Physical Kinetics}. Pergamon,
  London.

\bibitem[{Linker and Dieterich(1992)}]{Linker1992}
Linker, M.~F., Dieterich, J.~H., 1992. {Effects of variable normal stress on
  rock friction: Observations and constitutive equations}. Journal of
  Geophysical Research 97~(B4), 4923.

\bibitem[{Marone(1998)}]{Marone1998}
Marone, C.~J., 1998. {Laboratory-derived friction laws and their application to
  seismic faulting}. Annual Review of Earth and Planetary Sciences 26~(1),
  643--696.

\bibitem[{Marone and Scholz(1988)}]{Marone1988}
Marone, C.~J., Scholz, C.~H., 1988. {The depth of seismic faulting and the
  upper transition from stable to unstable slip regimes}. Geophysical Research
  Letters 15~(6), 621--624.

\bibitem[{Marone et~al.(1991)Marone, Scholz, and Bilham}]{Marone1991}
Marone, C.~J., Scholz, C.~H., Bilham, R., 1991. {On the mechanics of earthquake
  afterslip}. Journal of Geophysical Research 96~(B5), 8441.

\bibitem[{Martins et~al.(1995)Martins, Guimarães, and Faria}]{Martins1995b}
Martins, J., Guimarães, J., Faria, L.~O., 1995. {Dynamic surface solutions in
  linear elasticity and viscoelasticity With frictional boundary conditions}.

\bibitem[{Martins and Sim\~{o}es(1995)}]{Martins1995a}
Martins, J., Sim\~{o}es, F. M.~F., 1995. {On some sources of
  instability/ill-posedness in elasticity problems with Coulomb’s friction}.
  In: Raous, M., Jean, M., Moreau, J. (Eds.), Contact Mechanics. Springer US,
  pp. 95--106.

\bibitem[{Meziane et~al.(2007)Meziane, D'Errico, Baillet, and
  Laulagnet}]{Meziane2007}
Meziane, A., D'Errico, S., Baillet, L., Laulagnet, B., 2007. {Instabilities
  generated by friction in a pad-disc system during the braking process}.
  Tribology International 40~(7), 1127--1136.

\bibitem[{Nagata et~al.(2008)Nagata, Nakatani, and Yoshida}]{Nagata2008}
Nagata, K., Nakatani, M., Yoshida, S., 2008. {Monitoring frictional strength
  with acoustic wave transmission}. Geophysical Research Letters 35~(6),
  L06310.

\bibitem[{Nakatani(2001)}]{Nakatani2001}
Nakatani, M., 2001. {Conceptual and physical clarification of rate and state
  friction: Frictional sliding as a thermally activated rheology}. Journal of
  Geophysical Research 106~(B7), 13347--13380.

\bibitem[{Noda and Shimamoto(2009)}]{Noda2009}
Noda, H., Shimamoto, T., 2009. {Constitutive properties of clayey fault gouge
  from the Hanaore fault zone, southwest Japan}. Journal of Geophysical
  Research 114~(B4), B04409.

\bibitem[{Perfettini and Ampuero(2008)}]{Perfettini2008}
Perfettini, H., Ampuero, J.-P., 2008. {Dynamics of a velocity strengthening
  fault region: Implications for slow earthquakes and postseismic slip}.
  Journal of Geophysical Research 113~(B9), B09411.

\bibitem[{Prakash(1998)}]{Prakash1998}
Prakash, V., 1998. {Frictional response of sliding interfaces subjected to time
  varying normal pressures}. Journal of Tribology 120~(1), 97.

\bibitem[{Prakash and Clifton(1993)}]{Prakash1993}
Prakash, V., Clifton, R., 1993. {Time resolved dynamic friction measurememts in
  pressure-shear}. In: Ramesh, K. (Ed.), Experimental techniques in the
  dynamics of deformable solids. American Society of Mechanical Engineers.
  Applied Mechanics Division, pp. 33--48.

\bibitem[{Prakash and Clifton(1992)}]{Prakash1992}
Prakash, V., Clifton, R.~J., 1992. {Pressure-shear plate impact measurement of
  dynamic friction for high speed machining applications}. In: Proceedings of
  the 7th International Congress on Experimental Mechanics. pp. 8--11.

\bibitem[{Ranjith(2009)}]{Ranjith2009}
Ranjith, K., 2009. {Destabilization of long-wavelength Love and Stoneley waves
  in slow sliding}. International Journal of Solids and Structures 46~(16),
  3086--3092.

\bibitem[{Ranjith(2014)}]{Ranjith2014}
Ranjith, K., 2014. {Instabilities in dynamic anti-plane sliding of an elastic
  layer on a dissimilar elastic half-space}. Journal of Elasticity 115~(1),
  47--59.

\bibitem[{Ranjith and Rice(2001)}]{Ranjith2001}
Ranjith, K., Rice, J.~R., 2001. {Slip dynamics at an interface between
  dissimilar materials}. Journal of the Mechanics and Physics of Solids 49~(2),
  341--361.

\bibitem[{Renardy(1992)}]{Renardy1992}
Renardy, M., 1992. {Ill-posedness at the boundary for elastic solids sliding
  under Coulomb friction}. Journal of Elasticity 27~(3), 281--287.

\bibitem[{Rice(1993)}]{Rice1993}
Rice, J.~R., 1993. {Spatio-temporal complexity of slip on a fault}. Journal of
  Geophysical Research 98~(B6), 9885.

\bibitem[{Rice et~al.(2001)Rice, Lapusta, and Ranjith}]{Rice2001}
Rice, J.~R., Lapusta, N., Ranjith, K., 2001. {Rate and state dependent friction
  and the stability of sliding between elastically deformable solids}. Journal
  of the Mechanics and Physics of Solids 49~(9), 1865--1898.

\bibitem[{Rice and Ruina(1983)}]{Rice1983}
Rice, J.~R., Ruina, A., 1983. {Stability of steady frictional slipping}.
  Journal of applied mechanics 50, 343.

\bibitem[{Richardson and Marone(1999)}]{Richardson1999}
Richardson, E., Marone, C., 1999. {Effects of normal stress vibrations on
  frictional healing}. Journal of Geophysical Research 104~(B12), 28859.

\bibitem[{Ruina(1983)}]{Ruina1983}
Ruina, A., 1983. {Slip instability and state variable friction laws}. Journal
  of Geophysical Research 88~(B12), 10359--10370.

\bibitem[{Scholz(2002)}]{Scholz2002}
Scholz, C.~H., 2002. {The Mechanics of Earthquakes and Faulting}. Cambridge
  University Press.

\bibitem[{Shibazaki and Iio(2003)}]{Shibazaki2003}
Shibazaki, B., Iio, Y., 2003. {On the physical mechanism of silent slip events
  along the deeper part of the seismogenic zone}. Geophysical Research Letters
  30~(9), 1489.

\bibitem[{Sim\~{o}es and Martins(1998)}]{Simoes1998}
Sim\~{o}es, F. M.~F., Martins, J., 1998. {Instability and ill-posedness in some
  friction problems}. International Journal of Engineering Science 36~(11),
  1265--1293.

\bibitem[{Teufel and Logan(1978)}]{Teufel1978}
Teufel, L.~W., Logan, J.~M., 1978. {Effect of displacement rate on the real
  area of contact and temperatures generated during frictional sliding of
  Tennessee sandstone}. Pure and Applied Geophysics 116~(4-5), 840--865.

\bibitem[{Tonazzi et~al.(2013)Tonazzi, Massi, Culla, Baillet, Fregolent, and
  Berthier}]{Tonazzi2013}
Tonazzi, D., Massi, F., Culla, A., Baillet, L., Fregolent, A., Berthier, Y.,
  2013. {Instability scenarios between elastic media under frictional contact}.
  Mechanical Systems and Signal Processing 40~(2), 754--766.

\bibitem[{Weeks(1993)}]{Weeks1993}
Weeks, J.~D., 1993. {Constitutive laws for high-velocity frictional sliding and
  their influence on stress drop during unstable slip}. Journal of Geophysical
  Research 98~(B10), 17637.

\bibitem[{Weertman(1963)}]{Weertman1963}
Weertman, J., 1963. {Dislocations moving uniformly on the interface between
  isotropic media of different elastic properties}. Journal of the Mechanics
  and Physics of Solids 11~(3), 197--204.

\bibitem[{Weertman(1980)}]{Weertman1980}
Weertman, J., 1980. {Unstable slippage across a fault that separates elastic
  media of different elastic constants}. Journal of Geophysical Research
  85~(B3), 1455.

\bibitem[{Xia et~al.(2004)Xia, Rosakis, and Kanamori}]{Xia2004}
Xia, K., Rosakis, A.~J., Kanamori, H., 2004. {Laboratory earthquakes: the
  sub-Rayleigh-to-supershear rupture transition.} Science 303~(5665), 1859--61.

\end{thebibliography}

\end{document}